\documentclass[11pt]{article}
\begin{document}
\begin{center}
\today \\
\begin{Large} {\bf   Rotation in classical zero-point radiation
and in quantum vacuum. }
\end{Large}
\vspace{10mm}
\begin{it}
{Yefim S. Levin }\\
Department of Electrical and Computer Engineering, Boston University,
Boston MA, 02215   \\
\end{it}
\end{center}
\begin{abstract}
\indent Two reference systems, rotating $\{\mu_\tau\}$  and non
rotating $\{\lambda_\tau\}$, are defined and used as the basis for
investigating thermal effects of rotation through both random
\emph{classical} zero point radiation and \emph{quantum} vacuum.
Both reference systems consist of an infinite number of inertial
reference frames $\mu_\tau$ and $\lambda_\tau$ respectively.  The
$\mu$ and $\lambda$ reference frames do not accompany the detector
and are defined so that at each moment of proper time $\tau$ of
the detector there are two inertial frames, $\mu_\tau$ and
$\lambda_\tau$, which agree momentarily, are connected by a
Lorentz transformation with  the detector velocity as a parameter,
and with  origins at the detector location at the same time
$\tau$.\\
\indent   The two- field correlation functions measured by the
observer rotating through a random classical zero point radiation
, have been calculated and presented in terms of elementary
functions for both \emph{electromagnetic} and \emph{massless
scalar}  fields. \\
\indent If the correlation functions are periodic with a period
$\frac{2\pi}{\Omega}$ of rotation the observer finds the spectrum
which is very similar, but not identical, to Plank spectrum.\\
\indent  If both fields of  such a two-field periodic correlation
function, for both electromagnetic and massless scalar case, are
taken at the same point then its convergent (regularized ) part is
shown, using Abel-Plana summation formula, to have Planck spectrum
with the temperature $T_{rot}=\frac{\hbar \Omega}{2 \pi k}$. \\
\indent The convergent (regularized) part of the electromagnetic
energy density at the rotating detector is shown to have Plank
spectrum $reg \;w(\mu)=\frac{2(4 \gamma^2-1)}{3}\;w(T_{rot})$
where $w(T_{rot})=\frac{4\sigma}{c}T_{tot}^4$ is the energy
density of the black radiation at the temperature $T_{rot}$  and
the factor $\frac{2(4 \gamma^2-1)}{3}$ is a relativistic
anisotropy factor.\\
\indent It is shown that the vacuum of the quantized massless
scalar
 field in rotating  reference system $\{\mu_\tau\}$ is not equivalent
 to the vacuum
 of the field in the laboratory system because
 the respective Bogolubov transformation is not a zero.
\end{abstract}
\section{Introduction}.
 \indent The main issue related with a description of classical
 and quantum effects
 connected with rotation is a definition of a reference system.
 We will use   the concepts of laboratory coordinate system,
reference frames, and reference systems. \\
\indent The origin of the laboratory coordinate system is chosen
at the center of the rotating detector circle. At each proper time
of the rotating detector an inertial reference frame $\mu_\tau$
exists. The reference frame  $\mu_\tau$ is a global 3-dimensional
orthogonal system with constant vector velocity $\vec{v}_\tau$
relative to the laboratory coordinate system. The index $\tau$
means that $\mu_\tau$  is used by the detector only once,
momentarily, at the time $\tau$ and the origin of $\mu_\tau$ is
instantaneously at rest relative to the rotating detector and
coincides with it at the time $t_\tau=\tau$, where  $\vec{v}_\tau$
is the vector velocity of the detector at the proper time $\tau$,
and $t_\tau$ is the time measured in the reference frame.   The
first axis of the reference frame is directed along an
instantaneous radius of the detector $r_\tau $ in the laboratory
coordinate system. The second one is directed along the
instantaneous velocity vector $\vec{v}_\tau$  of the detector in
the laboratory system and the third one is perpendicular to the
plane of the detector rotation. Such reference frames can be
defined at each proper time $\tau$.
 The description of  zero-point vacuum
fluctuations in  the global reference frame  is  restricted in our
consideration
to a Minkovski space-time with pseudo-Cartesian coordinates.\\
\indent  A rotating reference system $\{\mu_\tau\}$is defined as a
set of
reference frames $\mu_\tau$ for all possible values of $\tau$.\\
 \indent A given RF, $\mu_\tau$,
is connected  by a  Lorentz transformation,  with  velocity
$|\vec{v}_\tau|=\Omega r$ depending on radius of rotation , to
another RF , $\lambda_\tau$. The index $\tau$ means again that
this RF $\lambda\tau$, its origin and axis directions , agrees
instantaneously with $\mu_\tau$ at detector proper time $\tau$.
Each RF $\lambda_\tau$ is a global one, at rest relative to the
laboratory coordinate system, and related with the laboratory
system by spatial shift and rotation transformations. A set of all
$\lambda_\tau$ defines a non rotating reference system
$\{\lambda_\tau\}$. Both systems are used to calculate the
correlation functions ( CF ) of the random classical
electromagnetic and massless scalar fields at the rotating
detector and Bogolubov transformations for a quantized massless
scalar field between $\{\mu_\tau\}$ and
the laboratory system.\\
\indent The correlation functions (CF) of random classical fields
measured by a uniformly accelerated detector have been
investigated and used in many works \cite{boyer1980, boyer1984,
heish1994, cole1986}. Our approach in CF calculation is very close
but not identical to the method developed in \cite{boyer1980,
boyer1984} for uniformly accelerated detectors . The reference
frames $\mu_\tau$ are similar to the frames $I_\tau$ defined in
\cite{boyer1984}, p.1091, but the reference frames $\lambda_\tau$
 are not used by Boyer
\cite{boyer1984}. The inertial frame $I_\ast$ agrees with the
frame $I_\tau$ (\emph{only}) at $\tau=0$ \cite{boyer1984}, p.
1091. We discuss this
issue in the Appendix B in detail.\\
\indent The case of the rotating detector, to the best of our
knowledge, has not been considered in random \emph{classical}
electromagnetic radiation. In \emph{quantum} case it was studied
for the massless scalar field in connection with rotating vacuum
puzzle \cite{davis1996, lorency2000, suga1999}, Bogolubov
transformation, and different coordinate mappings.  In this
article no special mapping between $(t,r,\theta, \phi)$ and
$(t^\prime,r^\prime,\theta^\prime, \phi^\prime)$ is used.
 The Bogolubov transformation between the
\emph{quantized} massless scalar fields defined in two different
reference systems is calculated based on our definitions of
$\{\mu_{\tau}\}$ and $\{\lambda_{\tau}\}$. It is done in Section
~\ref{sec-bogolubov}. The same reference systems are used for the
calculation of CF of a \emph{classical} zero-point electromagnetic
zero-point radiation and a \emph{classical} massless scalar field.
 \\
 \indent Reference system $\{\mu_\tau\}$  differs from the system of
 successive rest systems in constant circular motion introduced
 by Moller \cite{moller}, IV, \$47 because they have different
 initial conditions. This issue is also discussed in the Appendix
 B. \\
\indent  The calculation of the CF for an electromagnetic field is
presented in the Section ~\ref{sec-electromagnetic}.  The
calculation of  the CF for a massless scalar  field is performed
 in the Section ~\ref{sec-massless} in a similar manner, and it is
 much simpler than the calculation in the electromagnetic field case
because the scalar field does not change under Lorentz
transformations. This CF calculated in \emph{classical} approach
is identical with the correlation function of the rotating vacuum
of massless scalar field, obtained
 in the \emph{quantum} case \cite{davis1996}. \\
\indent  There is a simple relationship between proper time of the
rotating detector and the time measured in the lab system. Because
of that a period $T_\gamma =T/\gamma$ can be introduced, where
$T_\gamma$ is the time measured by the rotating detector and which
corresponds to the period of rotation, measured in the rotating
reference system $\mu_\tau$. We expect that the physical picture
of the vacuum observed by the rotating detector at two moments of
proper time separated by $T_\gamma$ is the same. For example,
$CF(\tau) = CF(\tau + T_\gamma)$. This assumption is investigated
below. The direct consequence of this periodicity condition is a
change in the spectrum of random electromagnetic field observed by
a rotating detector. It can measure the frequencies $\omega=
\Omega n$ only, where $n=0,\pm 1, \pm 2, \pm3,...$ and $\Omega$ is
an angular detector velocity. In the Section
~\ref{sec-classicalSpectrum} this periodicity condition is taken
into consideration, the final expression of the CF is modified,
and the integration over the absolute value of the wave vector is
changed to the summation over $k_n=k_0 n=\frac{\Omega}{c}n$. Using
Abel-Plana summation formula it is shown that the difference
between the infinite values of the energy density with the
discrete spectrum and of electromagnetic zero-point radiation with
the continuous spectrum is finite at a reference frame $\mu$ and
has the same spectrum as the Plank spectral function with
temperature $T_{rot}=\frac{\hbar \Omega}{2 \pi k}$. For the
radiation, observed by the rotating detector, the simple
relationship exists: $w(\mu)=w(T)\frac{4 \gamma^2-1}{3\pi}$ where
$w(T)$ is the Plank energy density radiation at the temperature T,
excluding zero-point radiation, and the factor $\frac{4
\gamma^2-1}{3 \pi}$ is a consequence of  the anisotropy of the
radiation observed by the moving relativistic detector.
\section{Electromagnetic field. Correlation functions in the case at a
rotating detector. \label{sec-electromagnetic} } \indent In this
section we will calculate the following two-field correlation
functions measured by a detector when it experiences a rotation
motion through the classical random zero-point radiation:
\\
$\langle E_i(\tau_2)E_j(\tau_1)\rangle$, $\langle
E_i(\tau_2)H_j(\tau_1)\rangle$, $\langle
H_i(\tau_2)H_j(\tau_1)\rangle$, \\
 where $i,j=1,2,3$ and electric and magnetic field intensities
 $E_i$, $H_i$  are measured by the  detector  at its proper
 times $\tau_1$ and $\tau_2$.   \\
\indent We follow the common idea \cite{boyer1984}, p.1091 that
all measurements at the time $\tau$ are carried out by non
inertial detector using an instantaneous inertial reference frame
which is momentarily at rest relative to the detector at $\tau$.
Therefore we  rewrite these expressions as
follows:\\
$\langle E_i(\mu_2|A_2^{\mu_2}, \tau_2)E_j(\mu_1|A_1^{\mu_1},
\tau_1)\rangle$, $\langle H_i(\mu_2|A_2^{\mu_2},
\tau_2)H_j(\mu_1|A_1^{\mu_1}, \tau_1)\rangle$, and $\langle
E_i(\mu_2|A_2^{\mu_2}, \tau_2)H_j(\mu_1|A_1^{\mu_1},
\tau_1)\rangle$, \\
So a two-field correlation function definition is based on two
measurements in the instantaneous  inertial reference frames
$\mu_1$ and $\mu_2$, with the detector at the points $A_1^{\mu_1}$
and $A_2^{\mu_2}$ of $\mu_1$ and $\mu_2$ at proper times $\tau_1$
and $\tau_2$ respectively. In each reference frame $\mu$, the time
is measured by the clock of the point detector. So $\tau_1=
t_1^{\mu_1}$ and $\tau_2=t_2^{\mu_2}$. The origin of the reference
frame $\mu_\tau$ is defined at the detector position at the proper
time $\tau$, so $A_1^{\mu_1}=0$, and $A_2^{\mu_2}=0$. But it is
convenient to keep $A_1^{\mu_1}$, and $A_2^{\mu_2}$  in this
unspecified form.\\
\indent To execute the operation of the averaging $\langle
\rangle$ of the expressions with two components of the
electromagnetic field, defined in two \emph{different} reference
frames, they should be transformed
 to a \emph{single} reference frame, for example $\lambda_{\tau_2}$
of the non rotating reference system. \\
\indent The main steps of the calculation of the CF $\langle
E_1(\mu_2|A_2^{\mu_2},
 \tau_2)E_1(\mu_1|A_1^{\mu_1},\tau_1)\rangle$are as follows
( the
same way all other CF's can be evaluated):\\
 1. Using Lorentz transformation,
$ E_1(\mu_2|A_2^{\mu_2},\tau_2)$ defined at the point
$A_2^{\mu_2}$ at the time $t^{\mu_2}=\tau_2$ in the $\mu_2$
reference frame is expressed in  terms of the electromagnetic
field components, defined at the point $A_2^{\lambda_2}$ at the
time $t_2^{\lambda_2}$  at the reference frame $\lambda_2$ .\\
2.$ E_1(\mu_1|A_1^{\mu_1},\tau_1)$, using again  Lorentz
transformation, is expressed in terms of the components at the
reference frame $\lambda_1$ at the point $A_1^{\lambda_1}$ at the
time $t_1^{\lambda_1}$. Of course the reference frames $\lambda_1$
and $\lambda_2$ have different directions and different origins in
the laboratory coordinate system. \\
3. Using rotation transformation, all the quantities in the
reference frame $\lambda_1$ are transformed to the  reference
frame $\lambda_1^\prime$ which, by definition, should have the
same axes
directions as $\lambda_2$.\\
 4. In the next step, a shift
transformation from $\lambda_1^\prime$ to $\lambda_2$ is made.
After  these steps, all random electromagnetic field intensities
in the CF are defined  in the inertial reference frame
$\lambda_2$. Expressions for random field  intensities in an
inertial reference frame are well known and given in
\cite{boyer1980}, formulae (48) and (49). \\
 5. Using the obtained expressions, the operation $\langle \rangle$
is executed in the reference frame $\lambda_2$.\\
 6. Summation over polarizations is made.\\
  7. Finally the CF is expressed as 3-dimensional integral in wave vector space.\\
   8. Its integrand  expression is
simplified using a rotation  transformation in ($k_1, k_2$) plane
of the wave vector space.
\subsection{Expressions for field components, defined in two different
reference frames, in terms of one reference frame.} \indent We
will first treat as an example with
\begin{eqnarray}
\langle E_1(\mu_1|A_1^{\mu_1}, \tau_1)E_1(\mu_2|A_2^{\mu_2},
\tau_2)\rangle    \label{eq: CF}
\end{eqnarray}
and will follow the directions described above.\\
\indent  The fluctuating electric and magnetic fields in an
inertial reference frame $\mu_\tau$ or $\lambda_\tau$ can be
written as \cite{boyer1980}
\begin{eqnarray}
\vec{E}(\vec{r},t)= \sum^2_{\lambda=1} \int d^3k
\hat{\epsilon}(\vec{k},\lambda)h_0(\omega)\cos[\vec{k}\vec{r}-\omega
t -\Theta(\vec{k},\lambda)], \nonumber \\
\vec{H}(\vec{r},t)=\sum^2_{\lambda=1} \int d^3k
[\hat{k},\hat{\epsilon}^(\vec{k},\lambda)]h_0(\omega)\cos[\vec{k}\vec{r}-\omega
t-\Theta(\vec{k},\lambda)]. \label{eq:ff}
\end{eqnarray}
where the $\theta(\vec{k},\lambda)$ are random phases distributed
uniformly on the interval $(0,2\pi)$ and independently for each
wave vector $\vec{k}$ and polarization $\lambda$ of of a plane
wave, and
\begin{eqnarray}
 \pi^2 h_0^2(\omega)=(1/2)\hbar \omega.
 \end{eqnarray}
\indent For special Lorentz transformation without rotation
between $\lambda_i$ and $\mu_i$, the transformation equations for
$\vec{E}$ and $\vec{H}$ at the points $A_1$ and $A_2$ can be
written in the form \cite{moller}, V.(15)
\begin{eqnarray}
\vec{E}(\mu_i|A_i^{\mu_i}, \tau_i)= \gamma
\vec{E}(\lambda_i|A_i^{\lambda_i},
t_i^{\lambda_i})+\frac{\vec{v}^{\lambda_i}}{v^2}(\vec{v}^{\lambda_i}
\vec{E}(\lambda_i|A_i^{\lambda_i},
t_i^{\lambda_i}))(1-\gamma)+\gamma  \frac{[\vec{v}^{\lambda_i},
\vec{H}(\lambda_i|A_i^{\lambda_i}, t_i^{\lambda_i}) ]}{c},& \\
\label{eq:Lorentz1} \vec{H}(\mu_i|A_i^{\mu_i}, \tau_i)= \gamma
\vec{H}(\lambda_i|A_i^{\lambda_i},t_i^{\lambda_i})
+\frac{\vec{v}^{\lambda_i}}{v^2}(\vec{v}^{\lambda_i}\vec{H}(\lambda_i|A_i^{\lambda_i},t_i^{\lambda_i})
)(1-\gamma)-\gamma \frac{[\vec{v}^{\lambda_i},
\vec{E}(\lambda_i|A_i^{\lambda_i},t_i^{\lambda_i})]}{c},
\label{eq:Lorentz2}
\end{eqnarray}
where $i=1,2$, $\gamma= \sqrt{1-\frac{v^2}{c^2}}$ and v is a
linear velocity of the detector. Its absolute value is constant,
and $ v=\Omega r$. Here $\Omega$ is an angular velocity of the
rotating detector. Vector $\vec{v}^{\lambda_i}$ ($i=1,2$  )is a
velocity vector of the inertial reference frame $\mu_i$  relative
to the inertial reference frame $\lambda_i$. \\
\indent Because $\vec{v}^\lambda = (v_1^\lambda, v_2^\lambda,
v_3^\lambda)= (0, v, 0)$ in both reference frames,  $\lambda_1$
and $\lambda_2$, we have:
\begin{eqnarray}
\label{eq:muTOlambda} E_1(\mu_i|A_i^{\mu_i},\tau_i)= \gamma
\emph{(} E_1(\lambda_i|A_i^{\lambda_i},
t_i^{\lambda_i})+\frac{v}{c}H_3(\lambda_i|A_i^{\lambda_i},t_i^{\lambda_i})
\emph{)}, \;\; i=1,2.
\end{eqnarray}
\indent For $i=1$ and $i=2$, the quantities on the right side of
these equations are still defined in different reference frames,
$\lambda_1$ and $\lambda_2$. Let us transform
$E_1(\lambda_1|A_1^{\lambda_1},\tau_1)$ and
$H_3(\lambda_1|A_1^{\lambda_1},\tau_1)$ in the last formula from
$\lambda_1$ to $\lambda_1^\prime$ (which, by definition,  has the
same axes directions as $\lambda_2$)
 by rotation  and then to $\lambda_2$ by
shifting. Then we have
\begin{eqnarray}
 E_1(\mu_1|A_1^{\mu_1},\tau_1)= \gamma
\emph{(} E_1(\lambda_1^\prime|A_1^{{\lambda_1}^\prime},
t_1^{{\lambda_1}^\prime})\cos\delta+
E_2(\lambda_1^\prime|A_1^{{\lambda_1}^\prime},
t_1^{{\lambda_1}^\prime})(-\sin\delta) + \frac{v}{c}
H_3(\lambda_1^\prime|
A_1^{{\lambda_1}^\prime}, t_1^{{\lambda_1}^\prime})\emph{)}\nonumber \\
\label{eq:muTOlambda1} = \gamma
\emph{(}E_1(\lambda_2|A_1^{\lambda_2}, t_1^{\lambda_2})\cos\delta+
E_2(\lambda_2|A_1^{\lambda_2}, t_1^{\lambda_2})(-\sin\delta) +
\frac{v}{c} H_3(\lambda_2| A_1^{\lambda_2},
t_1^{\lambda_2})\emph{)},
\end{eqnarray}
 where $\delta$ is an angle between
$\lambda_1$ and $\lambda_2$ references frames in the laboratory
coordinate system, and   $\delta=\Omega \gamma (\tau_2-\tau_1)$
.The explicit expressions for the coordinates of the points $A_1$
and $A_2$ in different reference frames that is $A_1^{\mu_1}$,
$A_1^{\lambda_1^{\prime}}$, and $A_1^{\lambda_2}$ are provided below.\\
\indent Taking into consideration (\ref{eq:muTOlambda}) and
(\ref{eq:muTOlambda1}), the CF can be written in the following
form
\begin{eqnarray}
 \langle E_1(\mu_1|A_1^{\mu_1},
\tau_1)E_1(\mu_2|A_2^{\mu_2}, \tau_2)\rangle & =
 \langle \gamma \; \textbf{[}
E_1(\lambda_2|A_1^{\lambda_2}, t_1^{\lambda_2})\cos\delta+
E_2(\lambda_2|A_1^{\lambda_2},
t_1^{\lambda_2})(-\sin\delta) + \nonumber \\
& +  \frac{v}{c} H_3(\lambda_2|A_1^{\lambda_2},
t_1^{\lambda_2}\textbf{]} \;
 \gamma \; \textbf{[} E_1(\lambda_2|A_2^{\lambda_2},
t_2^{\lambda_2})+
 \frac{v}{c}H_3(\lambda_2|A_2^{\lambda_2},t_2^{\lambda_2})
\textbf{]} \rangle.  \label{eq:CFinTermsOfOneFrame}
\end{eqnarray}
All field quantities on the right side of the equation are defined
in the same reference frame $\lambda_2$ and can be substituted by
the  expressions (\ref{eq:ff}). But first we have to specify all
arguments of these quantities, that is $A_1^{\lambda_2}$,
$A_2^{\lambda_2}$, $t_1^{\lambda_2}$, and $t_2^{\lambda_2}$.
\subsection{Coordinates of two detector locations in terms of one
reference frame.} \indent By definition of a $\mu_i$ reference
frame (we use $\mu_i$ instead of $\mu_{\tau_i}$ for simplicity
when such reference does not lead to confusion ), its origin
should be at the location of the detector and the detector proper
time $\tau_i$ should be equal to $t_i^{\mu_i}$ that is
\begin{eqnarray}
A_1^{\mu_1} =(x_1^{\mu_1}, x_2^{\mu_1}, x_3^{\mu_1})=(0,0,0), &&
t_1^{\mu_1}=\tau_1,    \\
A_2^{\mu_2} =(x_1^{\mu_2}, x_2^{\mu_2}, x_3^{\mu_2})=(0,0,0),
&& t_2^{\mu_2}=\tau_2.
\end{eqnarray}
The coordinates  of $\vec{A}^{\mu}$ and $\vec{A}^{\lambda}$ of the
point $\vec{A}$ in two reference frames $\mu$ and $\lambda$, are
connected with a Lorentz transformation without rotation and with
a special initial condition (see Appendix):
\begin{eqnarray}
\vec{x}^\lambda=\vec{x}^\mu + \vec{v}^\mu [\frac{\vec{x}^\mu
\vec{v}^\mu}{v^2}(\gamma -1)-t^\mu \gamma]+\vec{a}^\mu_\tau,
\end{eqnarray}
where
\begin{eqnarray}
\vec{a}^{\mu_1}_{\tau_1}=(0, v\tau_1 ,0)  &&
\vec{a}^{\mu_2}_{\tau_2}=(0, v\tau_2,0)
\end{eqnarray}
and $v$ and $\tau_i$ are parameters of the transformation.\\
 Then
\begin{eqnarray}
\vec{A}_1^{\lambda_1}=(x_1^{\lambda_1}, x_2^{\lambda_1},
x_3^{\lambda_1})=(0,0,0), \;\;\; t_1^{\lambda_1}=\gamma \tau_1,
\;\;\; \vec{A}_2^{\lambda_2}=(x_1^{\lambda_2}, x_2^{\lambda_2},
x_3^{\lambda_2})=(0,0,0), \;\;\; t_2^{\lambda_2}=\gamma \tau_2.
\label{eq:coordinate1}
\end{eqnarray}
\indent At each moment of proper time,$\tau_i$, the detector is at
the origin of both references frames $\lambda_i$ and $\mu_i$. \\
\indent After rotation from $\lambda_1$ to $\lambda_1^\prime$ and
shift from $\lambda_1^\prime$ to $\lambda_2$ the coordinates of
$A_1$ point are in $\lambda_1^\prime$ and $\lambda_2$ reference
frames respectively:
\begin{eqnarray}
 A_1^{\lambda_1^\prime}=(x_1^{\lambda_1^\prime},
x_2^{\lambda_1^\prime}, x_3^{\lambda_1^\prime})=(0,0,0),\;\;
t_1^{\lambda_1^\prime }=\gamma \tau_1, \;\;
A_1^{\lambda_2}=(-r(1-\cos \delta),-r \sin\delta,0), \;\;
t_1^{\lambda_2}=\gamma \tau_1, \label{eq:coordinate2}
\end{eqnarray}
where $ \delta=\Omega (t_2-t_1)=\Omega\gamma(\tau_2-\tau_1)$ is a
rotation angle of the detector for the time
$t_2-t_1$ .\\
 The last two
expressions will be used in the next subsection to get general
expression to calculate (\ref{eq:CFinTermsOfOneFrame}).
\subsection{General expression for the correlation function
$E_1(\mu_1|0,0,0,\tau_1)E_1(\mu_2|0,0,0,\tau_2) \rangle$.}
 Using
(\ref{eq:coordinate1}) and (\ref{eq:coordinate2}) expressions
(\ref{eq:muTOlambda}) for $i=2$ and (\ref{eq:muTOlambda1}) can be
written as
\begin{eqnarray}
E_1(\mu_1|A_1^{\mu_1},\tau_1)= \gamma (
E_1(\lambda_2|-r(1-\cos\delta),-r \sin\delta,0,\gamma \tau_1)
cos\delta  &  \nonumber \\+ E_2(\lambda_2| -r(1-\cos\delta),-r
\sin\delta,0,\gamma
\tau_1)(-\sin\delta) & \nonumber \\
+ \frac{v}{c} H_3(\lambda_2|-r(1-\cos\delta),-r \sin\delta,0,\gamma \tau_1
)),
\end{eqnarray}
and
\begin{eqnarray}
E_1(\mu_2|A_2^{\mu_2},\tau_2)= \gamma (
E_1(\lambda_2|,0,0,0,\gamma
\tau_2)+\frac{v}{c}H_3(\lambda_2|0,0,0,\gamma \tau_2) ).
\end{eqnarray}
\indent Each field intensity component on the right sides in these
expressions are defined in the single inertial reference frame
$\lambda_2$.
%
%
Having inserted them into (\ref{eq:CFinTermsOfOneFrame})and using
(\ref{eq:ff}) we arrive at the following expression for the CF
\begin{eqnarray}
\label{eq:CF_11} \langle
E_1(\mu_1|0,0,0,\tau_1)E_1(\mu_2|0,0,0,\tau_2) \rangle= \langle
\sum^{2}_{\lambda_1=1}\sum^{2}_{\lambda_2=1} \int d^3k_1 \int
d^3k_2
h_0(\omega_1)h_0(\omega_2) \gamma^2 \times          \nonumber     \\
\{\hat{\epsilon}_{1x}(\vec{k}_1\lambda_1)\cos\delta +
\hat{\epsilon}_{1y}(\vec{k}_1\lambda_1)(-\sin \delta)+
(\hat{k}_{1x} \hat{\epsilon}_{1y} (\vec{k}_1 \lambda_1)-
\hat{k}_{1y} \hat{\epsilon}_{1x} (\vec{k}_1 \lambda_1)
) \frac{v}{c} \}\times                                          \nonumber  \\
\{\hat{\epsilon}_{2x}(\vec{k}_2\lambda_2) +
(\hat{k}_{2x}\hat{\epsilon}_{2y}(\vec{k}_2 \lambda_2)-
\hat{k}_{2y}\hat{\epsilon}_{2x}(\vec{k}_2 \lambda_2))
\frac{v}{c}\} \times
\nonumber \\
\cos\{ k_{1x} [-r(1-\cos\delta)] +k_{1y} (-r \sin\delta) -\omega_1
\gamma \tau_1
 -\theta(\vec{k_1}\lambda_1)\}  \cos\{-\omega_2\gamma\tau_2 -\theta(\vec{k_2}
\lambda_2)\}\rangle,
\end{eqnarray}
where  symbols $\lambda_1$ and $\lambda_2$ are polarizations, not
reference frame labels.\\
\indent Taking into consideration
\cite{boyer1980} that
\begin{eqnarray}
\label{eq:coseverage0} \langle \cos\theta(\vec{k}_1 \lambda_1
)\cos\theta(\vec{k}_2 \lambda_2 ) \rangle=\langle
\sin\theta(\vec{k}_1 \lambda_1 )\sin\theta(\vec{k}_2 \lambda_2 )
\rangle=\frac{1}{2}\delta_{\lambda_1
\lambda_2}\delta^3(\vec{k}_1-\vec{k}_2)
\end{eqnarray}
and
\begin{eqnarray}
\label{eq:polarization}
\sum^2_{\lambda=1}\epsilon_i(\vec{k}\lambda
)\epsilon_j(\vec{k}\lambda )= \delta_{ij}-k_ik_j / k^2
\end{eqnarray}
and after integrating over $\vec{k}_1$ and summing over
$\lambda_1$ this expression can be reduced to
\begin{eqnarray}
 \langle
E_1(\mu_1|0,0,0,\tau_1)E_1(\mu_2|0,0,0,\tau_2)\rangle =
\int d^3k h^2_0(\omega) \gamma^2 \frac{1}{2} \times \nonumber \\
\emph{(}\cos\delta - \hat{k}_x\frac{v}{c}\sin\delta -\hat{k}_y 2
\frac{v}{c}\cos^2 \frac{\delta}{2} + \hat{k}_x\hat{k}_y\sin\delta
+ \hat{k}^2_x (-\cos\delta + \frac{v^2}{c^2}) +\hat{k}^2_y\frac
{v^2}{c^2}\emph{)}\times \nonumber  \\
 \cos \{ r [k_x(1-\cos\delta)+k_y \sin \delta ]-\omega \gamma(
\tau_2-\tau_1 ) \},
\end{eqnarray}
where
\begin{eqnarray}
\hat{k_x}=\frac{k_x}{k},& \hat{k_y}=\frac{k_y}{k}, &
\hat{k_z}=\frac{k_z}{k}.
\end{eqnarray}
The integrand of the integral can be simplified by the variable
change:
\begin{eqnarray}
\label{eq:wavevectors}
  \hat{k}_x^{\prime}=\hat{k}_x\cos \frac {\delta}{2}
-\hat{k}_y\sin\frac{\delta}{2}  \nonumber  \\
\hat{k}_y^{\prime}=\hat{k}_x\sin
\frac{\delta}{2}+\hat{k}_y\cos\frac{\delta}{2}.
\end{eqnarray}
The terms which are odd in $k_x$  vanish. Finally the correlation
function takes the form:
\begin{eqnarray}
\label{eq:CF_E_11} \langle
E_1(\mu_1|0,0,0,\tau_1)E_1(\mu_2|0,0,0,\tau_2)\rangle =
\int d^3k h^2_0(\omega) \gamma^2 \frac{1}{2} \times \nonumber \\
\emph{(}\cos\delta -\hat{k}_y 2 \frac{v}{c}\cos \frac{\delta}{2} +
+ \hat{k}^2_x (-\cos^2 \frac{\delta}{2} + \frac{v^2}{c^2})
+\hat{k}^2_y(\sin^2 \frac{\delta}{2}+ \frac
{v^2}{c^2})\emph{)}\times \nonumber  \\
\cos\emph{(} 2kr\sin\frac{\delta}{2}\hat{k}_y -c k(t_2 -t_1)
\emph{)}.
\end{eqnarray}
We have omitted primes in this expression.\\
\subsection{General expressions for other correlation functions.}
\indent Similar expressions can be obtained for other CFs:
\begin{eqnarray}
\langle E_2(\mu_1|A_1^{\mu_1}, \tau_1)\:E_2(\mu_2|A_2^{\mu_2},
\tau_2)\rangle=\langle
E_2(\mu_1|0,0,0,\tau_1)\:E_2(\mu_1|0,0,0,\tau_2) \rangle = \nonumber \\
\int d^3k \: h^2_0(\omega)\: \frac{1}{2} \times \: \frac{1}{2}[ \:
\hat{k}^2_x- \hat{k}^2_y \: + \: ( \:1+ \hat{k}^2_z \:)
 \cos \delta ]\: \times
\cos\emph{(} 2kr\sin\frac{\delta}{2}\hat{k}_y -c k(t_2 -t_1)
\emph{)}.
\end{eqnarray}
The third diagonal element of the electrical part of the CF is
\begin{eqnarray}
\langle E_3(\mu_1|A_1^{\mu_1}, \tau_1)\:E_3(\mu_2|A_2^{\mu_2},
\tau_2)\rangle= \langle
E_3(\mu_1|0,0,0,\tau_1)\:E_3(\mu_1|0,0,0,\tau_2) \rangle = \nonumber \\
\int d^3k \: h^2_0(\omega)\: \frac{1}{2} \times \gamma^2 \{ \: 1 +
\frac{v^2}{c^2} \cos \delta + \hat{k}_y (-2 \frac{v}{c} \cos
\frac{\delta}{2}) + \frac{v^2}{c^2}(-\hat{k}^2_x \cos^2
\frac{\delta}{2} + \hat{k}^2_y \sin^2 \frac {\delta}{2})
-\hat{k}^2_z
 \}\: \times \nonumber \\
\cos\emph{(} 2kr\sin\frac{\delta}{2}\hat{k}_y -c k(t_2 -t_1)
\emph{)}.
\end{eqnarray}
The non-diagonal elements of the electrical components of the CFs
are as follows:
\begin{eqnarray}
\langle E_1(\mu_1|A_1^{\mu_1}, \tau_1)\:E_2(\mu_2|A_2^{\mu_2},
\tau_2)\rangle=\langle
E_1(\mu_1|0,0,0,\tau_1)\:E_2(\mu_1|0,0,0,\tau_2) \rangle = \nonumber \\
\int d^3k \: h^2_0(\omega)\: \frac{1}{2} \: \times \{\:
 -(1+ \hat{k}^2_z)\: \frac{\gamma}{2} \sin \delta  + \hat{k}_y \: \gamma \frac {v}{c}
\sin \frac {\delta}{2} \}\: \times \cos\emph{(}
2kr\sin\frac{\delta}{2}\hat{k}_y -c k(t_2 -t_1) \emph{)}.
\end{eqnarray}
It is easy to see that
\begin{eqnarray}
\langle E_1(\mu_1|0,0,0,\tau_1)\:E_2(\mu_1|0,0,0,\tau_2) \rangle =
- \langle E_2(\mu_1|0,0,0,\tau_1)\:E_1(\mu_1|0,0,0,\tau_2)
\rangle.
\end{eqnarray}
The other non-diagonal elements of the CF with electric field
components are zeroes :
\begin{eqnarray}
\langle E_1(\mu_1|0,0,0,\tau_1)\:E_3(\mu_1|0,0,0,\tau_2) \rangle =
\langle E_3(\mu_1|0,0,0,\tau_1)\:E_1(\mu_1|0,0,0,\tau_2) \rangle = \nonumber \\
\langle E_2(\mu_1|0,0,0,\tau_1)\:E_3(\mu_1|0,0,0,\tau_2) \rangle =
\langle E_3(\mu_3|0,0,0,\tau_1)\:E_2(\mu_1|0,0,0,\tau_2) \rangle
=\emph{0}.
\end{eqnarray}
The CFs with magnetic components  are as follows:
\begin{eqnarray}
\langle H_1(\mu_1|A_1^{\mu_1}, \tau_1)\:H_1(\mu_2|A_2^{\mu_2},
\tau_2)\rangle=\langle
H_1(\mu_1|0,0,0,\tau_1)\:H_1(\mu_1|0,0,0,\tau_2) \rangle = \nonumber \\
\int d^3k \: h^2_0(\omega)\: \frac{1}{2} \: \times \gamma^2 \times
\{\: \hat{k}_y(\cos^2 \delta -2 \frac{v}{c}\cos \frac{\delta}{2})
+ \hat{k}_x^2 (-\frac{1}{2}\sin^2 \delta) + \hat{k}_y^2
(\frac{1}{2} \sin^2 \delta) + \hat{k}_z^2 \cos \delta) + \nonumber
\\ \hat{k}_y\hat{k}_z^2(-\cos\frac{\delta}{2}\cos \delta) +
\hat{k}_x^2\hat{k}_z^2 \sin^2\frac{\delta}{2}\cos\delta +
\hat{k}_y^2 \hat{k}_z^2 \cos^2\frac{\delta}{2} \cos\delta
 \} \times \nonumber \\
  \cos\emph{(}
2kr\sin\frac{\delta}{2}\hat{k}_y -c k(t_2 -t_1) \emph{)}.
\end{eqnarray}
So all CFs can be given as 3-dimensional integrals over $(k,
\theta, \phi)$. In the next subsection an example of calculation
of  these integrals is given.
\subsection{Integral calculations: final expression
for $E_1(\mu_1|0,0,0,\tau_1)E_1(\mu_2|0,0,0,\tau_2) \rangle$.}
\indent All non zero expressions for CFs  have a common integral
over k. It can be easily calculated:
\begin{eqnarray}
\label{eq:IntegralOverK}
 \int_0^\infty d k k^3 \cos \{ k ( 2 r \sin
 \frac{\delta}{2} \sin \theta \sin \phi -c (t_2 -t_1))\} =  \frac
{6}{ {\{2 r \sin \frac{\delta}{2}  \sin \theta \sin \phi -
c(t_2-t_1)\}^4}}= \nonumber \\
= \frac {6}{[ c(t_2-t_1)]^4} \frac{1}{ [1-\frac{v}{c}\frac{\sin
\delta/2}{\delta/2}\sin\theta \sin \phi]^4 }.
\end{eqnarray}
The integrals over $\theta$ and $\phi$ can be represented  in
terms of elementary functions. Let us show it  for $\langle
E_1(\mu_1|0,0,0,\tau_1)E_1(\mu_1|0,0,0,\tau_2) \rangle$
\begin{eqnarray}
\langle E_1(\mu_1|0,0,0,\tau_1)E_1(\mu_1|0,0,0,\tau_2) \rangle =
  \frac{3\hbar c}{2 \pi^2
[c(t_2-t_1)]^4} \gamma^2 \int_{0}^{\pi} d\theta  \nonumber \\
\times \{(\cos \delta \sin \theta +(-\cos^2 \frac{\delta}{2}+
\frac{v^2}{c^2}) \sin^3 \theta) \int_{0}^{2\pi} d\phi \:
\frac{1}{(1+b \sin\phi)^4} \nonumber \\
+ (-2\frac{v}{c}\cos\frac{\delta}{2})sin^2\theta\int_{0}^{2\pi}
d\phi \: \frac{\sin\phi}{(1+b \sin\phi)^4}
+\sin^3\theta\int_{0}^{2\pi} d\phi \: \frac{\sin^2\phi}{(1+b
\sin\phi)^4} \},
\end{eqnarray}
where $b \equiv k \: \sin\theta, \;\; k \equiv
-\frac{v}{c}\frac{\sin\delta/2}{\delta/2}$. So   k is a constant,
not a wave vector.  \\
We have used here:
\begin{eqnarray}
\label{eq:HatVector} \hat{k}_x = \sin\theta \cos \phi ,  &
\hat{k}_y= \sin\theta \sin\phi ,  & \hat{k}_z= \cos \theta.
\end{eqnarray}
\indent The next step is to calculate the integral over $\phi$.
Because \cite{gr1965}
\begin{eqnarray}
\int_0^{2\pi}d\phi \frac{1}{(1+b \sin \phi)^4}=\frac{\pi(2 + 3
b^2)}{(1-b^2)^{7/2}},
\end{eqnarray}
\begin{eqnarray}
\int_0^{2\pi}d\phi \frac{\sin \phi}{(1+b \sin \phi)^4}=
\frac{-b\pi (4+b^2)}{(1-b^2)^{7/2}},
\end{eqnarray}
and
\begin{eqnarray}
\int_0^{2\pi}d\phi \frac{\sin^2 \phi}{(1+b \sin \phi)^4}=
\frac{\pi(1+4b^2)}{(1-b^2)^{7/2}},
\end{eqnarray}
the correlation function takes the form:
\begin{eqnarray}
\langle E_1(\mu_1|0,0,0,\tau_1)E_1(\mu_1|0,0,0,\tau_2) \rangle =
\frac{3\hbar c}{2 \pi^2 [c(t_2-t_1)]^4} \gamma^2 \{ + [2 \pi \cos
\delta] \int_{0}^{\pi}d\theta\frac{\sin\theta}
{(1-k^2\sin^2\theta)^{7/2}} \nonumber \\
+[3\pi k^2 \cos\delta -2 \pi \cos^2(\delta/2)+ 2\pi \beta^2 - 8
\pi \beta k \cos(\delta/2)+\pi]
\int_{0}^{\pi}d\theta\frac{\sin^3\theta}
{(1-k^2\sin^2\theta)^{7/2}} \nonumber \\
+[-3 \pi k^2 \cos^2(\delta/2) + 3 \pi \beta^2 k^2 -2 \pi \beta k^3
\cos(\delta/2) + 4 \pi k^2]
\int_{0}^{\pi}d\theta\frac{\sin^5\theta}
{(1-k^2\sin^2\theta)^{7/2}} \},
\end{eqnarray}
where (\cite{prudnikov}, 1.5.23, 1.2.43:
\begin{eqnarray}
\int_{0}^{\pi}d\theta\frac{\sin\theta}
{(1-k^2\sin^2\theta)^{7/2}}= \frac{2}{5(1-k^2)} +
\frac{8}{15(1-k^2)^2} + \frac{16}{15(1-k^2)^3} , \label{eq:sin1} \\
 \int_{0}^{\pi}d\theta\frac{\sin^3\theta}
{(1-k^2\sin^2\theta)^{7/2}}= \frac{4}{15(1-k^2)^2}
+\frac{16}{15(1-k^2)^3}, \label{eq:sin3}\\
\int_{0}^{\pi}d\theta\frac{\sin^5\theta}
{(1-k^2\sin^2\theta)^{7/2}}=\frac{16}{15(1-k^2)^3}.
\label{eq:sin5}
\end{eqnarray}
\subsection{The Features of the $E_1(\mu_1|0,0,0,\tau_1)E_1(\mu_2|0,0,0,\tau_2) \rangle$.}
 The correlation function depends on  the difference
 $|\tau_2 - \tau_1|$,
 a reasonable property of
 a correlation function, and parameters $\Omega$,
 $\beta=\frac{\Omega r}{c}$, and k. The parameter k depends on
 $\delta$ according to
$k=-\frac{v}{c}\frac{\sin(\delta/2)}{\delta/2}$. The $\delta$ is
the angle the detector has rotated for
the time $t_2-t_1$. \\
\indent  If $\delta \rightarrow 0$  then $k=-\beta$, and
\begin{eqnarray}
\langle E_1(\mu_1|0,0,0,\tau_1)E_1(\mu_1|0,0,0,\tau_1 \pm 0)
\rangle = \frac{3\hbar c}{2 \pi^2 [c(t_2-t_1)]^4} \gamma^2 \{ +2
\pi[\frac{2}{5(1-\beta^2)} + \frac{8}{15(1-\beta^2)^2}+
\frac{16}{15(1-\beta^2)^3}]  \nonumber \\
 + (-3 \pi
\beta^2 -\pi)[\frac{4}{15(1-\beta^2)^2} +
\frac{16}{15(1-\beta^2)^3}]+ (\pi\beta^2 + \pi
\beta^4)\frac{16}{15(1-\beta^2)^3} \}.
\end{eqnarray}
When $\beta \rightarrow 0$ then the value of this function is
\begin{eqnarray}
\langle E_1(\mu_1|0,0,0,\tau_1)E_1(\mu_1|0,0,0,\tau_1 \pm 0)
\rangle_{\beta \rightarrow 0} = \frac{4 \hbar
c}{\pi[c(t_2-t_1)]^4},
\end{eqnarray}
and it is positively defined as it is supposed to be for the
quantity which is  a contribution to the energy density of the
electromagnetic field. The function is divergent when $t_1
\rightarrow t_2$. \\
\indent The expressions (\ref{eq:sin1}) and (\ref{eq:sin3}) have
been obtained based on the \cite{prudnikov}, 1.2.43, 1.5.23 and
(\ref{eq:sin5}). \\
\indent In the next section we will show that the CFs should be
modified to take its periodicity into consideration.
\section{The Spectrum of the random classical zero-point
electromagnetic radiation observed by a rotating detector.}
\label{sec-classicalSpectrum}
\subsection{Periodicity of the correlation function  and Abel-Plana
formula. }
 Later we will see that in the quantum case the Bogolubov
 coefficients are periodic due to the periodic motion of the
 rotating detector. We can expect that in a classical case
periodic motion of a rotating detector should also result in
periodicity of its measurements and particularly in the
correlation function. Mathematically it means  that
 $\langle E_1(\mu_1|0,t_1)E_1(\mu_2|0,t_2)
\rangle=\langle E_1(\mu_1|0,t_1)E_1(\mu_{2n}|0,t_2+
\frac{2\pi}{\Omega}n) \rangle$.  Here $\Omega=\frac{2\pi}{T}$ is
an angular velocity of the rotating detector and $n= \pm 0, 1, 2,
3,...$. It is easy to show that (\ref{eq:CF_E_11}) is periodic
 if $\omega=\Omega n$.
 It means that the rotating detector observes the random
 electromagnetic radiation with the same
discrete spectrum  as a rotating electrical charge radiates
\cite{ivanenko}(39.29).  Let us
consider other consequences of the periodicity.\\
\indent The equations (\ref{eq:ff}) for the discrete spectrum
should be
 modified to:
\begin{eqnarray}
\vec{E}(\vec{r},t)= a\:\sum^{\infty}_{n=0} \sum^2_{\lambda=1} \int
do\,k^2_n\,\hat{\epsilon}(\hat{k},\lambda)\,h_0(\omega_n)\
\cos[\vec{k_n}\vec{r}-\omega_n
t -\Theta(\vec{k_n},\lambda)], \nonumber \\
\vec{H}(\vec{r},t)=a\:\sum^{\infty}_{n=0}\sum^2_{\lambda=1} \int
do \,k^2_n \,
[\hat{k},\hat{\epsilon}(\hat{k},\lambda)]\,h_0(\omega_n)\,\cos[\vec{k_n}\vec{r}-\omega_n
t-\Theta(\vec{k_n},\lambda)], \nonumber \\
 \vec{k}_n=k_n \hat{k},
\;\; k_n=k_0\,n, \;\; k_0= \frac{\Omega}{c}, \;\; \omega_n=c
\,k_n, \;\;
 do= d\theta \, d\phi \, \sin\theta, \nonumber \\
 \hat{k}=(\hat{k}_x,
\hat{k}_y, \hat{k}_z)=(\sin\theta \, \cos \phi, \, \sin\theta \,
\sin\phi, \,\cos\theta\,), \;\; a=c \Omega. \label{eq:mff}
\end{eqnarray}
 The unit vector $\hat{k}$ defines a
direction of the wave vector and does not depend on its value,
n.\\
The correlation function (\ref{eq:CF_11})  takes the form:
\begin{eqnarray}
\label{eq:CF_11discrete} \langle
E_1(\mu_1|0,0,0,\tau_1)E_1(\mu_2|0,0,0,\tau_2) \rangle= a^2
\langle \;\;\sum^{\infty}_{n_1, n_2=0}\;\;
 \sum^{2}_{\lambda_1,\lambda_2=1} \int do_1 \; do_2 \;
h_0(\omega_{n_1}) \, h_0(\omega_{n_2}) \gamma^2 \times          \nonumber     \\
\{\hat{\epsilon}_{1x}(\hat{k}_1\lambda_1)\cos\delta +
\hat{\epsilon}_{1y}(\hat{k}_1\lambda_1)(-\sin \delta)+
(\hat{k}_{1x} \hat{\epsilon}_{1y} (\hat{k}_1 \lambda_1)-
\hat{k}_{1y} \hat{\epsilon}_{1x} (\hat{k}_1 \lambda_1)
) \frac{v}{c} \}\times                                          \nonumber  \\
\{\hat{\epsilon}_{2x}(\hat{k}_2\lambda_2) +
(\hat{k}_{2x}\hat{\epsilon}_{2y}(\hat{k}_2 \lambda_2)-
\hat{k}_{2y}\hat{\epsilon}_{2x}(\hat{k}_2 \lambda_2))
\frac{v}{c}\} \times
\nonumber \\
\cos\{ k_{n_1} \,[\hat{k_{1x}} [-r(1-\cos\delta)] +\hat{k_{1y}}
(-r \sin\delta) -c \gamma \tau_1]
 -\theta(\hat{k_1}\lambda_1)\}  \cos\{-k_{n_2}c \,
 \gamma\tau_2 -\theta(\vec{k_2}
\lambda_2)\}\rangle.
\end{eqnarray}
In spherical coordinates the right side of the relation
(\ref{eq:coseverage0}) should be changed \cite{davydov1968},
p.656:
\begin{eqnarray}
\label{eq:cosaverage} \langle \cos\theta(\vec{k}_1 \lambda_1
)\cos\theta(\vec{k}_2 \lambda_2 ) \rangle=\langle
\sin\theta(\vec{k}_1 \lambda_1 )\sin\theta(\vec{k}_2 \lambda_2 )
\rangle=\frac{1}{2}\delta_{\lambda_1
\lambda_2}\delta^3(\vec{k}_1-\vec{k}_2)=
\frac{1}{2}\delta_{\lambda_1 \, \lambda_2} \, \frac{2}{k_1^2}\,
\delta(k_1-k_2)\delta(\hat{k}_1-\hat{k}_2).
\end{eqnarray}
In the case of the discrete spectrum it  takes the form:
\begin{eqnarray}
\label{eq:cosaverage} \langle \cos\theta(\vec{k}_{n_1} \lambda_1
)\cos\theta(\vec{k}_{n_2} \lambda_2 ) \rangle= \langle
\sin\theta(\vec{k}_{n_1} \lambda_1 )\sin\theta(\vec{k}_{n_2}
\lambda_2 )\rangle=  \frac{1}{2}\delta_{\lambda_1 \, \lambda_2} \,
\frac{2}{k_0(k_0 n_1)^2}\, \delta_{n_1 \,
n_2}\delta(\hat{k}_1-\hat{k}_2).
\end{eqnarray}
The equation (\ref{eq:polarization})
 $\;\; \sum^2_{\lambda=1}\epsilon_i(\vec{k}\lambda
)\epsilon_j(\vec{k}\lambda )= \delta_{ij}-\hat{k}_i\hat{k}_j \;\;$
does not depend on n. \\
Then the correlation function (\ref{eq:CF_11discrete})  becomes
\begin{eqnarray}
\label{eq:CF discrete} \langle
E_1(\mu_1|0,0,0,\tau_1)E_1(\mu_2|0,0,0,\tau_2)\rangle =a^2 \int do
 \gamma^2 \{\cos\delta -\hat{k}_y  \frac{2v}{c}\cos
\frac{\delta}{2} +  \hat{k}^2_x (-\cos^2 \frac{\delta}{2} +
\frac{v^2}{c^2}) +\hat{k}^2_y(\sin^2 \frac{\delta}{2}+ \frac
{v^2}{c^2})\} \nonumber  \\
\times \sum_{n=0}^\infty(k_0 \, n)^2 \, h^2_0(c k_0 n)\, \cos[
k_0n \, (\,2r\sin\frac{\delta}{2}\hat{k}_y -c(t_2 -t_1)\,) ]=
\nonumber
\\
=\frac{a^2 \, \gamma^2 \,k_0^3 c \hbar}{2 \pi ^2}   \int \, do \,
\emph{[}\cos\delta -\hat{k}_y 2 \frac{v}{c}\cos \frac{\delta}{2} +
\hat{k}^2_x (-\cos^2 \frac{\delta}{2} + \frac{v^2}{c^2})
+\hat{k}^2_y(\sin^2 \frac{\delta}{2}+ \frac
{v^2}{c^2})\emph{]}\times \sum_{n=0}^\infty  n^3 \, \cos n\,F,
\end{eqnarray}
where
\begin{eqnarray}
\label{eq:F}
 F= k_0 \, (\,2r\sin\frac{\delta}{2}\hat{k}_y -c(t_2
-t_1)\,)=
 \delta\,[ 1 - \frac{v}{c}\frac{\sin
\delta/2}{\delta/2}\sin\theta \sin \phi],\;\;
\delta=\Omega(t_2-t_1).
\end{eqnarray}
The sum over n in  this equation
\begin{eqnarray}
S \equiv
 \sum_{n=0}^\infty n^3 \cos (n \, F)
\end{eqnarray}
can be evaluated using the Abel-Plana summation formula
\cite{Bateman1953},\cite{T1988}, \cite{Evgrafov1968}:
\begin{eqnarray}
\label{eq:AbelPlana}
 \sum_{n=0}^\infty \, f(n)= \int_0^\infty
f(x)\,dx + \frac{f(0)}{2} +i \,\int_0^\infty \,dt \,
\frac{f(it)-f(-it)}{e^{2 \pi t}-1},
\end{eqnarray}
 Having utilized this formula and following to
\cite{MT1988} we come to the following expression:
\begin{eqnarray}
\label{eq:sum}
 \Omega^4 S
 =  \int _0^{\infty} d \, \omega \omega^3 \cos( \omega \tilde{F})+
 \int_0^{\infty} d\omega \frac{2 \omega^3 \cosh(\omega \tilde{F})}{e^
 {2\pi \omega/\Omega}-1}, & \tilde{F}=\frac{F}{\Omega}.
\end{eqnarray}
The integrals in this expression can be computed and put in  two
forms:
\begin{eqnarray}
\label{eq:I_one}
 S = \frac{6}{F^4} + [\frac{3 -2 \sin^2(F/2)}{8
\sin^4(F/2)} -\frac{6}{F^4}]
\end{eqnarray}
or
\begin{eqnarray}
\label{eq:I_two}
 S= \frac{6}{F^4} +
6  \sum_{n=1}^{\infty} \frac{1}{(2\pi n)^4} [
\frac{1}{(1+\frac{F}{(2\pi n)^4})}  + \frac{1}{(1-\frac{F}{(2\pi
n)^4})} ].
\end{eqnarray}
Complete calculation of the integrals over $\theta$ and $\phi$ in
the (\ref{eq:CF discrete}) will be given elsewhere. In this
article we wold like to focus on the expression (\ref{eq:sum})
which allows a simple physical interpretation. It will be done in
the next subsection.

\subsection{Planck spectrum of the energy density of random classical
electromagnetic radiation observed by a rotating detector.}
 Let us compare the expression
(\ref{eq:sum})for $\emph{S}$ with the  expression \cite{boyer1980}
(74) for the Fourie component of the spectral function
$\frac{1}{2}\hbar \omega \coth \frac{\hbar \omega}{2 k T}$ of the
electromagnetic radiation with Planck's spectrum at the
temperature T,with  the zero-point radiation:
\begin{eqnarray}
\label{eq:Tspectral function} \frac{1}{2}\int_0^{\infty} d\omega
\omega^3 \coth(\frac{\hbar \omega}{2 k T})\cos \omega t=
\frac{1}{2}\;[\;\int_0^{\infty}d\omega\omega^3 \cos \omega t +
\int_0^{\infty}d \omega \frac{2 \omega ^3}{e^{\frac{\hbar
\omega}{k T}-1}} \cos \omega t\;].
\end{eqnarray}
The right sides of these expressions are very similar, except for
two features:   in the first expression \emph{$\tilde{F}$} and
\emph{cosh}  are used instead of \emph{t} and \emph{cos}
respectively in (\ref{eq:Tspectral function})
 . For $\tilde F=0$ and $t=0$ though the right sides of both
expressions are identical if we define a new variable $T_{rot}$
according to:
\begin{equation}
\label{eq:T}
 T_{rot}=\frac{\hbar \Omega}{2 \pi k},
\end{equation}
where k is a Bolzman constant.\\
\indent This remarkable resemblance brings up the idea that the
energy density of the random classical electromagnetic radiation
measured by a detector, rotating through a zero point radiation,
has the Planck spectrum at the temperature $T_{rot}$ (\ref{eq:T}).
\\
\indent Using the  technique, described above for a discrete
spectrum, the energy density
\begin{equation}\label{eq:energy density}
w(\mu)=\frac{1}{8 \pi}\langle \sum_{i=1}^3(E_i^2(\mu|)+
H_i^2(\mu|))\rangle=\frac{1}{4\pi}\{[E_1^2(\lambda)+
E_3^2(\lambda)]\gamma^2(1+\beta^2)+E_2^2(\lambda)\},
\end{equation}
measured by the rotating observer at an instantaneous inertial
reference frame $\mu$, can be given in the form:
\begin{equation}
\label{eq: energy density 1}
w(\mu)= \frac{ 2(4\gamma^2-1)}{3 }\;\frac{\hbar}{c^3 \pi^2}\;
\frac{1}{2}\Omega^4\sum_{n=0}^\infty n^3
\end{equation}
or with the help of (\ref{eq:sum}) for $F=0$ as
\begin{equation}
\label{eq:energy density 2} w(\mu)= \frac{2(4\gamma^2-1)}{3 } \;
w(T_{rot}),
\end{equation}
where
\begin{eqnarray}
\label{eq: energy density 3}
w(T_{rot})=\frac{\hbar }{c^3 \pi^2}
\;\frac{1}{2}( \; \int _0^{\infty} d \, \omega \omega^3 +
 \int_0^{\infty} d\omega \frac{2 \omega^3 }
 {e^{\hbar\omega/kT_{rot}} -1}
\;).
\end{eqnarray}
 is the full (with a zero point radiation included ) averaged
energy density measured by an inertial observer at the temperature
$T_{rot}$. The $w(\mu)$ does not depend on time and we omitted the
index t (or $\tau$). In the limiting case of $\Omega \rightarrow
0$, $T_{rot}\rightarrow 0$, $\gamma \rightarrow 1$, and $\langle
w(\mu)\rangle = \langle w(T_{rot}=0)\rangle$. In the relationship
(\ref{eq:energy density 2})  $\langle w(T_{rot})$ is divergent for
any $T_{rot}$.
\\
\indent The energy density (\ref{eq: energy density 3})has two
terms and it is the first term which  is divergent. The second one
is connected with periodicity of the detector rotation. It is
convergent. Usually such convergent term is  referred to  as a
regularized energy density $reg \; w(\mu)$ [\cite{MT1988}], p.969
and considered as an observable  physical quantity. It is equal to
\begin{eqnarray}
reg \: w(\mu) = \frac{2(4\gamma^2-1)}{3}\; w_{rad},
\end{eqnarray}
where
\begin{eqnarray}
w_{rad} = 4 \; \frac{\pi^2 k^4}{60 (c \hbar)^3}\; T_{rot}^4.
\end{eqnarray}
It is well known expression of the energy density of the black
radiation at the temperature $T_{rot}$ [\cite{landau_lifshits},
(60,14)]:
\begin{eqnarray}
 w_{rad}=\frac{4 \sigma}{c} T_{rot}^4,
 \end{eqnarray}
 and k is a Bolzman
constant, $\sigma$ is a Stephan-Bolzman constant , and $w_{rad}$
is the density of the energy of black radiation, without zero
point radiation,  at the
temperature $T_{rot}$.\\
\indent So, due to periodicity of the motion, an observer rotating
through a zero point radiation should see the energy density,
which would have been observed by an observer  moving in a thermal
bath at the temperature $T_{rot}=\frac{\hbar \Omega}{2 \pi k}$,
and multiplied by the factor $\frac{2}{3}(4\gamma^2-1)$. This
factor comes from integration in (\ref{eq:CF discrete}) over
angles and therefore is a consequence of anisotropy of the
electromagnetic field measured by an observer with velocity
$\beta$. When the angular velocity of the detector is zero, the
regularized energy density is zero as well. For a fixed angular
velocity, the energy density depends on a radius of the detector
circular path via $\gamma^2$.  When \emph{r} and therefore
$\gamma$   increases the regularized energy density increases as well. \\
\section{ Massless scalar field. Correlation function at a rotating
detector. } \label{sec-massless}
\subsection{Classical consideration.}
\indent The calculation of the correlation function for a
massless scalar field is
 much simpler than the calculation in the electromagnetic field case
because the scalar field does not change under Lorentz
transformations. The  correlation function measured by an observer
rotating through a classical massless zero-point  scalar field
radiation has the form:
\begin{eqnarray}
\langle \psi_s(\mu_1|A_1^{\mu_1}, t_1^{\mu_1}) \psi_s(\mu_2 |
A_2^{\mu_2}, t_2^{\mu_2})\rangle =\langle
\psi_s(\lambda_1|A_1^{\lambda_1}, t_1^{\lambda_1})
\psi_s(\lambda_2 |A_2^{\lambda_2},t_2^{\lambda_2}) \rangle = \nonumber \\
\langle \psi_s(\lambda_1^\prime|A_1^{\lambda_1^\prime},
t_1^{\lambda_1^\prime}) \psi_s(\lambda_2
|A_2^{\lambda_2},t_2^{\lambda_2}) \rangle = \langle
\psi_s(\lambda_2|A_1^{\lambda_2}, t_1^{\lambda_2})
\psi_s(\lambda_2 | A_2^{\lambda_2},t_2^{\lambda_2})\rangle,
\label{eq:CF_Scal_Field_Continuous}
\end{eqnarray}
where ($A_1^{\mu_1},t_1^{\mu_1}$),
($A_1^{\lambda_1},t_1^{\lambda_1}$),
($A_1^{\lambda_1^\prime},t_1^{\lambda_1^\prime}$), and
($A_1^{\lambda_2}, t_1^{\lambda_2}$) are 4-coordinates  of the
rotating detector at the first position, taken in the reference
frames $\mu_1$, $\lambda_1$, $\lambda_1^\prime$, and $\lambda_2$
respectively. Transitions between these reference frames were
discussed in a previous section. In the last expression of the
equation, all coordinates are defined in the same reference frame
$\lambda_2$. Then taking into consideration (\ref{eq:coordinate2})
we can write:
\begin{eqnarray}
\psi_s(\lambda_2 | A_1^{\lambda_2}, t_1^{\lambda_2})= \int d^3k_1
f(\omega_1) \cos \{ -k_{1x}r (1 - \cos \delta) -k_{1y} r \sin \delta -
\omega_1 \tau_1 \gamma - \theta(k_1)\}
\end{eqnarray}
\begin{eqnarray}
\psi_s(\lambda_2 | A_2^{\lambda_2}, t_2^{\lambda_2})=
\int d^3k_2 f(\omega_2) \cos \{- \omega_2 \tau_2 \gamma
- \theta(k_2)
\}
\end{eqnarray}
Using these expressions
 and \cite{boyer1980}
\begin{eqnarray}
 \langle \cos \theta(\vec{k}_1) \cos \theta (\vec {k}_2) \rangle
= \langle \sin \theta(\vec{k}_1) \sin \theta (\vec {k}_2) \rangle
= \frac{1}{2} \delta^3(\vec{k}_1 - \vec{k}_2), \;\;
f^2(\omega)=\frac{\hbar c^2}{2 \pi^2 \omega},
\end{eqnarray}
we come to the expression:
\begin{eqnarray}
\langle \psi_s(\mu_1|A_1^{\mu_1}, t_1^{\mu_1}) \psi_s(\mu_2 |
A_2^{\mu_2}, t_2^{\mu_2})\rangle=
 \int
d^3 k f^2(\omega)\frac{1}{2} \cos\{r (k_x (1 -\cos \delta) + k_y
\sin \delta)-\omega \gamma (\tau_2 - \tau_1)\}
\end{eqnarray}
or, after coordinate change (\ref{eq:wavevectors}) in the
integrand and omitting primes,  to:
\begin{eqnarray}
 \langle \psi_s(\mu_1|0,0,0,
\tau_1) \psi_s(\mu_2 |0,0,0, \tau_2)\rangle=
 \int d^3 k \times
\frac{\hbar c^2}{2 \pi^2 \omega}\times \frac{1}{2} \times \cos (2
r k_y \sin \frac {\delta}{2}-c k \gamma(\tau_2 - \tau_1)).
\end{eqnarray}
Having integrated it over k, the right side  takes the form:
\begin{eqnarray}
-\frac{\hbar c}{4\pi^2}\:\int_0^\pi d\theta \:\sin\theta \:
\int_0^{2 \pi} d\phi \: \: [E \:\sin \phi- B]^{-2}.
\end{eqnarray}
where $B=\gamma \tau c$, $E=2 r \sin\theta \: \sin \frac{ \Omega
\gamma \tau}{2}$, and $\tau=\tau_2- \tau_1$. \\
Because $B-|E|= c \gamma \tau \{1-\frac{v}{c} |\sin \theta \:
\frac{\sin \pi (\gamma \tau /T)}{\pi(\gamma \tau / T)}|\}> c
\gamma \tau (1 - v/c) >0,$
 and using \cite{prudnikov}
 we obtain :
\begin{eqnarray}
\int_0^{2 \pi} d\phi \: \frac {1}{[E \:\sin \phi- B]^2}= \frac{2
\pi B}{(B^2 - E^2)^{3/2}}.
\end{eqnarray}
 Having integrated over $\theta$ we come to the final expression
 of the CF for the correlation function of the random classical
 massless
 scalar field at the rotating  detector moving through a zero point
 fluctuating massless scalar radiation:
\begin{eqnarray}
\langle \psi_s(\mu_1|0,0,0, \tau_1) \psi_s(\mu_2 |0,0,0,
\tau_2)\rangle=
  -\frac{\hbar
c}{\pi}\frac{1}{(\gamma (\tau_2 - \tau_1)c)^2- 4 r^2 \sin^2
\frac{\Omega \gamma (\tau_2 - \tau_1)}{2}}.
\end{eqnarray}
\indent This correlation function received in the classical
approach based on two references systems $\mu_\tau$ and
$\lambda_\tau$  is identical, up to a constant factor, to the
Wightman function \cite{davis1996}, \cite{birell1982}(3.59)
received in the quantum case. \\
\indent The physical sense of  this function and its Fourie
component has been investigated by several authors in the frame of
a quantum theory, starting from Pfautsch[\cite{pfautsch1981}].
Davis, Dray, and Manogue [\cite{davis1996}] think that the
spectrum found in [\cite{pfautsch1981}] numerically is only "a
reminiscent of a Planck spectrum" because "the Bogolubov
transformation between rotating and non rotating modes is trivial"
and two "sets of modes are identical".  Recently De Lorency, De
Paola, and Svaiter [\cite{lorency2000}] using a proper mapping
between rotating and non rotating coordinate systems found new
modes of the scalar filed in the rotating system and showed that
the Bogolubov transformation is not zero. But this mapping has
very unusual features. So the question about non zero Bogolubov
transformation is still open for further
investigation.\\
\indent Below we investigate this issue again both in classical,
with some natural periodicity condition,  and quantum approach
calculating the Bogolubov transformation. Our consideration is
based on two reference systems and does not use mapping between
rotating and nonrotating coordinate systems.
\subsection{Quantum consideration. Bogolubov transformation between
modes of a massless scalar field in a rotating reference system
and the laboratory coordinate system.} \label{sec-bogolubov}
 \indent Usually the question if the
vacuums of massless scalar field observed by a rotating observer
and inertial one are unitary equivalent is discussed using
Bogolubov transformations between inertial ( laboratory )
reference system and a rotating reference system.  The history of
this issue is given  in \cite{lorency2000}. It was found that the
Bogolubov coefficients are null if rotating and inertial
coordinates are mapped as
\begin{eqnarray}
t=t^\prime, \;\; r=r^\prime, \;\; \theta=\theta^\prime -\Omega
t^\prime, z=z^\prime. \nonumber
\end{eqnarray}
The significant  feature of this coordinate system is that an
attendant rotating reference frame is co-moving with the observer.
In the attendant rotating reference frame the observer
is permanently at rest.\\
\indent Recently Lorenci, de Paola, and Svaiter \cite{lorency2000}
have shown that Trocheries -Takeno coordinates with non linear
connection between a linear and angular velocities of the observer
\begin{eqnarray}
t=t^\prime \cosh\Omega r^\prime -r^\prime \theta^\prime \sinh
\Omega r^\prime, \;\; r=r^\prime, \;\; \theta=\theta^\prime
\cosh\Omega r^\prime -\frac{t^\prime}{r^\prime}\sinh\Omega
r^\prime,\;\; z=z^\prime     \nonumber
\end{eqnarray}
 should be used to get non zero
Bogolubov coefficients. \\
In this case the location of the observer, $(R_0, \theta^\prime,
z^\prime=0)$ is constant in the rotating reference frame. But the
observer and the reference frame are not co-moving because the
metrics of the rotating reference frame depends on time and a
distance between any  point and the observer changes in time as
well. The rotating reference frame used in
\cite{lorency2000} is not a rigid one in the sense defined in
\cite{moller}. \\
 \indent The rotating reference system $\{\mu_t\}$ defined in
this work is also not co-moving with the rotating detector. It
consists of infinite number of inertial reference frames $\mu_t$
moving in the flat space-time of the laboratory system. They do
not accompany the observer. The frame $\mu_t$ labelled by t agrees
with the observer only once , instantaneously, at the respective
moment of time t. No special coordinates are used in these frames
and they have the Minkovsky metric.  The reference system
$\{\mu_t\}$ can be used for both classical and quantum systems. We
will see that the reference system $\{\mu_\tau\}$ is very useful
to calculate Bogolubov's coefficients. The final expression of the
coefficients is given in terms of elementary functions. It is much
simpler than the expression received in  \cite{lorency2000} and
explicitly not zero.\\
\indent Let us consider the quantized scalar massless scalar field
in two reference frames $\mu_t$ and $\lambda$.  The $\lambda$
reference frame  agrees with
$\mu_t$ with t=0. \\
\indent In both global inertial reference frames, $\mu_t$ and
$\lambda$, the Fourier series for the operators of scalar field
have  similar forms \cite{BjorDrell}
\begin{eqnarray}
\psi(\vec{x},t)=\int
d^3k[a(k)f_k(\vec{x},t)+a^{+}(k)f^{\ast}(\vec{x},t)], \nonumber \\
\psi^{\mu_t}(\vec{\xi^{\mu_t}},\eta^{\mu_t})=\int d^3k[a^{\mu_t}_k
f^{\mu_t}_k(\vec{\xi^{\mu_t}},\eta^{\mu_t}) + a^{\mu_t +}f^{\mu_t
\ast}_k(\vec{\xi^{\mu_t}},\eta^{\mu_t})].
\end{eqnarray}\\
For simplicity purposes we have omitted the index $\lambda$ in the
first expression. The operator $\psi(\vec{x},t)$ describes the
quantized scalar field in the laboratory inertial reference system
at the time t. The operator $\psi^{\mu_t}()$ describes the same
quantized scalar field in the inertial reference frame $\mu_t$
which is instantaneously at rest relative  to the detector at the
\emph{same} laboratory time t. There is a close relationship
between the coordinates $\vec{x},t$ in the laboratory reference
system $\lambda$ and $\vec{\xi}^{\mu_t}, \eta^{\mu_t}$ of the
inertial reference frame $\mu_t$. It is defined below. The
development of the $\psi$ operator in time in the laboratory
system corresponds to the description of the operator
$\psi^{\mu_t}$ in a \emph{system} of inertial reference frames
$\{\mu_t\}$, \emph{one  instantaneous
reference frame for each moment of time.}   \\
\indent The plane  waves have similar forms
\cite{BjorDrell},(9.6):
\begin{eqnarray}
\label{eq:plane waves}
 f_k(x, t)=\frac{1}{((2 \,
\pi)^3\, 2 \omega_k)^{1/2}}\exp\{-i \, ( \omega_k \, t -
\,\vec{k}\,\vec{x}
)\,\}, \nonumber \\
f^{\mu_t}_k(\,\vec{\xi}^{\mu_t},\,
\eta^{\mu_t}\,)=\frac{1}{(\,(2\, \pi)^3\, 2\,
\omega_k\,)^{1/2}}\exp\{-i (\, \omega_k \, \eta^{\mu_t} -
\vec{k}\,\vec{\xi}^{\mu_t} \, )\}.
\end{eqnarray}
 We assume that  operators $a(k)$ and $a^{\mu_t}$ are different in
 both reference frames and have to find the connection between
them. For scalar field we have:
\begin{eqnarray}
\label{eq:equalScalars0}  \psi(\vec{x}, t)=
 \psi^{\mu_t}(\vec{\xi}^{\mu_t},\,
\eta^{\mu_t}
 )
\end{eqnarray}
Here  spatial coordinates $ \vec{\xi}^{\mu_t}$ and time coordinate
$\eta^{\mu_t}$ are considered as functions of $\vec{x}$ and $t$.
In our case:
\begin{eqnarray}
\label{eq:ksiEtaVariables} \xi^{\mu_t}_1=x_1 \cos\delta_t + x_2
\sin\delta_t - 2 r
\sin^2\frac{\delta_t}{2}, \nonumber  \\
\xi^{\mu_t}_2= x_1 (-\gamma \sin \delta_t ) +x_2 \gamma \cos
\delta_t + t (-v \gamma)- (r \sin \delta_t + a_{\tau})  \gamma, \nonumber \\
\xi^{\mu_t}_3 = x_3, \nonumber \\
\eta^{\mu_t} = x_1 (\frac{v}{c^2} \gamma \sin \delta_t) + x_2
(-\frac{v}{c^2} \gamma \cos \delta_t) + t \gamma + (r \sin
\delta_t + a_{\tau})\frac{v}{c^2}\gamma.
\end{eqnarray}
These relationships have been obtained by three consequent
transformations. The first is a Lorentz transformation between
$\mu_t$ reference frame and $\lambda_t$. It was described above in
terms of $\mu_{\tau}$ and $\lambda_{\tau}$. The second one is a
rotation from $\lambda_t$ to $\lambda^{\prime}_t$. The
$\lambda^{\prime}_t$ reference frame is parallel to the $\lambda$
reference frame. And the third one is a shift from from
$\lambda^{\prime}_t$ to $\lambda$.This shift is an opposite to the
direction of the detector rotation. So $\psi^{\mu_t}$ depends on t
in two ways and such dependence on time finally makes the
Bogolubov's coefficients non zero. First, the variables
$\vec{\xi}^{\mu_t}$ and $\eta^{\mu_t}$ depend on t via the Lorentz
transformation . And second, the rotation by the angle
$\delta_t=\Omega t$, the parameter $a_\tau$ of the Lorentz
transformation, and the shift depend on t as well.  \\
\indent Having constructed  scalar products of
(\ref{eq:equalScalars0}) with $f_{k^\prime}$
\begin{eqnarray}
\int
d^3k[a(k)(f_{k^\prime},f_k)+a^{+}(k)(f_{k^\prime},f^{\ast}_k)]
=\int d^3k[a^{\mu_t}_k (f_{k^\prime},f^{\mu_t}_k) + a^{\mu_t
+}(f_{k^\prime},f^{\mu_t \ast}_k)]
\end{eqnarray}
and $f_{k^\prime}^{\ast}$
\begin{eqnarray}
\int
d^3k[a(k)(f^{\ast}_{k^\prime},f_k)+a^{+}(k)(f^{\ast}_{k^\prime},f^{\ast}_k)]
=\int d^3k[a^{\mu_t}_k (f^{\ast}_{k^\prime},f^{\mu_t}_k) +
a^{\mu_t +}(f^{\ast}_{k^\prime},f^{\mu_t \ast}_k)],
\end{eqnarray}
where a scalar product in the scalar field is defined according to
\cite{birell1982}, (2.9) [03.25.06, 37]:
\begin{eqnarray}
(f_{k^{\prime}},f^{ \mu_t}_k)=
 -i \int d^3
[f_{k^{\prime}}\frac{\partial (f^{ \mu_t})^{\ast}}{\partial t}
-\frac{\partial f_{k^{\prime}}}{\partial t}(f^{ \mu_t})^{\ast} ]
\end{eqnarray}
and
\begin{eqnarray}
(f_{k^\prime},f_k)=\delta^3(\vec{k}-\vec{k^\prime}),\;\;
f^{\ast}_{k^\prime},f^{\ast}_k)=-\delta^3(\vec{k}-\vec{k^\prime}),\;\;
f^{\ast}_{k^\prime},f_k)=f_{k^\prime},f^{\ast}_k)=0,
\end{eqnarray}
we arrive at the following relationships between $a_k$ and
$a^{\mu^t}_k$:
\begin{eqnarray}
\label{eq: creation_operators} a_{k^\prime} = \int d^3 k\;
[a_k^{\mu_t} \alpha^{\ast}_{kk
^\prime}+ a_k^{+ \mu_t}\beta_{kk^\prime}], \nonumber \\
a^{+}_{k^\prime} = \int d^3 k\; [a_k^{\mu_t} \beta^{\ast}_{kk
^\prime}+ a_k^{+ \mu_t}\alpha_{kk^\prime}].
\end{eqnarray}
 \indent  The Bogolubov coefficients $\beta_{kk^{\prime}}$ and
 $\alpha_{kk^\prime}$ are defined in
 our notations as follows:
\begin{eqnarray}
\beta_{kk^{\prime}}=(f_{k^{\prime}},f^{\ast \mu_t}_k), \nonumber \\
\alpha_{kk^\prime}=-(f^{\ast}_{k^\prime},f^{\ast \mu_t}_k).
\end{eqnarray}
\indent The most interesting for us is $\beta_{kk^\prime}$. Taking
into consideration (\ref{eq:plane waves}) $\beta_{kk^\prime}$ can
be given in the form:
\begin{eqnarray}
\beta_{k k^{\prime}}= -i \int d^3 x f_{k^{\prime}} f^{\mu_t}_k [-i
\omega_k \frac{\partial \eta^{\mu_t}}{\partial t} + i
\vec{k}\frac{\partial \vec{\xi}^{\mu_t}}{\partial t} +i
\omega_{k^{\prime}}].
\end{eqnarray}
Using (\ref{eq:ksiEtaVariables}) and
\begin{eqnarray}
\frac {\partial a_{\tau}}{\partial t} =-v, \nonumber \\
\frac{\partial \delta_t}{\partial t}=\Omega,
\end{eqnarray}
this expression finally equals to:
\begin{eqnarray}
\beta_{k k^{\prime}}= B_0 B_1 \int^{\infty}_{-\infty} dx_1 x_1
\exp(i x_1 A_1) \int^{\infty}_{-\infty} dx_2\exp(i x_2 A_2)
\int^{\infty}_{-\infty} dx_3\exp(i x_3 A_3) + \nonumber \\
+ B_0 B_2 \int^{\infty}_{-\infty} dx_1  \exp(i x_1 A_1)
\int^{\infty}_{-\infty} dx_2 x_2 \exp(i x_2 A_2)
\int^{\infty}_{-\infty} dx_3\exp(i x_3 A_3) + \nonumber \\
+ B_0 B_3 \int^{\infty}_{-\infty} dx_1  \exp(i x_1 A_1)
\int^{\infty}_{-\infty} dx_2\exp(i x_2 A_2)
\int^{\infty}_{-\infty} dx_3\exp(i x_3 A_3),
\end{eqnarray}
where
\begin{eqnarray}
B_0= \frac{-i}{(2 \pi)^3 2 (\omega_k
\omega_{k^{\prime}})^{1/2}}\exp\{-i t [\omega_{k^{\prime}}+
\frac{\omega_k} {\gamma}] +i r[- k_1 + k_1 \cos \delta_t -
(k_2 + \frac{v}{c^2}\omega_k)\gamma \sin \delta_t] \}, \nonumber \\
A_1=k^{\prime}_1 +k_1 \cos \delta_t - ( k_2 +
\omega_k \frac{v}{c^2}) \gamma \sin \delta_t , \nonumber \\
A_2=k^{\prime}_2 + k_1 \sin\delta_t +(k_2 + \omega_k
\frac{v}{c^2})\gamma \cos\delta_t, \nonumber \\
A_3= k^{\prime}_3 + k_3, \nonumber \\
B_1= -i \Omega [k_1 \sin \delta_t  + (k_2+ \omega_k
\frac{v}{c^2})\gamma \cos \delta_t
], \nonumber \\
B_2= -i \Omega [ k_1 \cos \delta_t -(k_2+ \omega_k
\frac{v}{c^2})\gamma \sin
\delta_t ], \nonumber \\
B_3= -i r\Omega [k_1 \sin\delta_t + (k_2 + \omega_k
\frac{v}{c^2})\gamma \cos \delta_t  ] + i(\omega_{k^{\prime}} -
\frac{\omega_k}{\gamma}).
\end{eqnarray}
\indent It is easy to show that the first two lines in the
expression for $\beta_{k k^{\prime}}$ are zeros because
\begin{eqnarray}
\int ^{+\infty}_{-\infty}dx x \exp\{i x A\}=0.
\end{eqnarray}
Indeed,  if $A=0$ then $\int ^{+\infty}_{-\infty}dx x =0$. If $A
\neq 0$
 then
\begin{eqnarray}
\int^{+\infty}_{-\infty}d x x \exp i x A = \lim_{\lambda
\rightarrow 0} \{\int^{\infty}_{0}d x x \exp  x(i A - \lambda ) -
\int^{\infty}_{0}d x x \exp - x (i A + \lambda \} = \nonumber \\
\lim_{\lambda \rightarrow 0}\frac{(2 \lambda)(2 i A)}{\lambda^2 +
A^2} =0. \nonumber
\end{eqnarray}
Besides
\begin{eqnarray}
\int d x_i \exp(i x_i A_i)=2 \pi \delta (A_i), \;\; i=1, 2, 3.
\nonumber
\end{eqnarray}
Then we obtain:
\begin{eqnarray}
\label{eq: beta} \beta_{k k^{\prime}}= \frac{-i}{2(\omega_k
\omega_{k^{\prime}})^{1/2}}\exp\{-i t [\omega_{k^{\prime}}+
\frac{\omega_k} {\gamma}] +i r[- k_1 - k_1^{\prime}]\}\; [-ir
\Omega(-k_2^{\prime})+i(\omega_{k^{\prime}}-
\frac{\omega_k}{\gamma})]  \times \nonumber \\
\times \;\delta \;(k_1^{\prime} + k_1 \cos \delta_t - (k_2 +
\omega_k \frac{v}{c^2}) \gamma \sin \delta_t) \;\; \delta \;
(k_2^{\prime} + k_1 \sin \delta_t + (k_2 + \omega_k \frac{v}{c^2})
\gamma \cos \delta_t)\;\; \delta\;(k_3^{\prime}+ k_3)
\end{eqnarray}
We have taken into consideration that, because of the
$\delta$-function features,  $\beta_{k k^{\prime}} \neq 0$ only if
\begin{eqnarray}
-k_1^{\prime} = k_1 \cos \delta_t - (k_2 + \omega_k \frac{v}{c^2})
\gamma \sin \delta_t \nonumber
\end{eqnarray}
and
\begin{eqnarray}
-k_2^{\prime} = k_1 \sin \delta_t + (k_2 + \omega_k \frac{v}{c^2})
\gamma \cos \delta_t \nonumber
\end{eqnarray}
\indent To interpret the physical sense of $\beta_{k k^{\prime}}$
let us evaluate the expectation value of the operator of number of
particles $N_{k^\prime}=a^{+}_{k\prime}a_{k^\prime}$ in the vacuum
state $|0^{\mu_t}>$  that is $<0^{\mu_t}|N_{k^\prime}|0^{\mu_t}>$.
Obviously that $<0|N_{k^\prime}|0>=0$ because $a_k|0>=0$ and
$<0|a_k^{+}=0$ by definition of the vacuum state $|0>$. It is easy
to see using (\ref{eq: creation_operators}) that
\begin{eqnarray}
<0^{\mu_t}|N_{k^\prime}|0^{\mu_t}>= \int d^3 \tilde{k}
\beta^{\ast}_{\tilde{k}k^\prime}\;<0^{\mu_t}|a_{\tilde{k}}^{\mu_t}
\;\;| \int d^3 k \beta_{kk^\prime} a^{+\mu_t}_k |0^{\mu_t}>
\end{eqnarray}
because $a^{\mu_t}_k|0^{\mu_t}>=0$ and
$<0^{\mu_t}|a_k^{+\mu_t}=0$. We can say that the vacuum
$|0^{\mu_t}>$ of modes $f_k^{\mu_t}$ contains
$<0^{\mu_t}|N_{k^\prime}|0^{\mu_t}>$ particles of modes
$f_k$ \cite{birell1982} (3.42). \\
\indent  We need to compute the integral operator with the
operator $a^{+\mu_t}(k)$ in its integrand
\begin{eqnarray}
\hat{I}_{k^\prime}=\int d^3 k \;\beta_{k k^{\prime}}\;
a^{+\mu_t}(k)=\int \frac{d^3k }{\omega_k}\;\omega_k \;\beta_{k
k^{\prime}}\;a^{+\mu_t}(k).
\end{eqnarray}
We use here   $a^{+\mu_t}(k)$ rather
than $a^{+\mu_t}_k$.\\
\indent To calculate the integral we have to change integrand
variables  to simplify arguments of the $\delta$ functions in
(\ref{eq: beta}). Let us use first the Lorentz transformation
between ($\omega_k, k_1, k_2, k_3$) and ($\omega_\kappa, \kappa_1,
\kappa_2, \kappa_3$):
\begin{eqnarray}
\omega_k=\gamma (\omega_{\kappa} - v \kappa_2), \;\; k_2=
\gamma(\kappa_2-\omega_{\kappa}\frac{v}{c^2}), \;\; k_1 =\kappa_1
\;\; k_3 = \kappa_3
\end{eqnarray}
or
\begin{eqnarray}
\omega_{\kappa}=\gamma(\omega_k + vk_2), \;\; \kappa_2=\gamma(k_2
+ \omega_k \frac{v}{c^2}), \;\; \kappa_1=k_1, \;\; \kappa_3=k_3.
\end{eqnarray}
Then the integral $\hat{I}_{k^{\prime}}$ takes the form:
\begin{eqnarray}
\hat{I}_{k^\prime}=\int \frac{d^3 \kappa }{\omega_{\kappa}}\;\;
\frac{[ \gamma ( \omega_{\kappa} - v \kappa_2 )]^{1/2}\;\;[v \;
k_2^{\prime}+(\omega_{k^{\prime}}- \frac{\gamma(\omega_{\kappa}-v
\kappa_2)}{\gamma})]} {2( \omega_{k^{\prime}} )^{1/2}}
\;\; G \times \nonumber \\
\delta(k_1^{\prime}+ \kappa_1 \cos \delta_t -\kappa_2
\sin\delta_t)\;\; \delta(k_2^{\prime} + \kappa_1 \sin \delta_t +
\kappa_2 \cos \delta_t)\;\; \delta(k_3^{\prime}+\kappa_3)\;
|a^{+\mu_t}(\kappa),
\end{eqnarray}
where
\begin{eqnarray}
 G=\exp\{-i t [\omega_{k^{\prime}}+ \frac{\gamma (\omega_\kappa
-v \kappa_2)} {\gamma}] +i r[- \kappa_1 - k_1^{\prime}]\}
\end{eqnarray}
 and we have used relationship
\begin{eqnarray}
\frac{d^3 k}{\omega_k}=\frac{d^3 \kappa}{\omega_{\kappa}},
\end{eqnarray}
which is a Lorentz transformation invariant \cite{Landau},
$\$10$.\\
 \indent The next variable changes in the integrand is the following
 rotation transformation  between ($\omega_\kappa, \kappa_1,
 \kappa_2,\kappa_3$) and ($\omega_k, k_1, k_2, k_3$):
\begin{eqnarray}
k_1 =\kappa_1 \cos \delta_t - \kappa_2 \sin \delta_t ,\;\; k_2 =
\kappa_1 \sin \delta_t + \kappa_2 \cos \delta_t , \;\; k_3=
\kappa_3, \;\; \omega_k=\omega_\kappa,
\end{eqnarray}
or
\begin{eqnarray}
\kappa_1 = k_1 \cos \delta_t + k_2 \sin\delta_t, \;\;
\kappa_2=-k_1\sin\delta_t + k_2 \cos \delta_t.
\end{eqnarray}
In these variables the integral $\hat{I}_{k^\prime}$ and its
integrand take the form:
\begin{eqnarray}
\hat{I}_{k^\prime}=  \int \frac{d^3k}{\omega_k}\;\; \frac{\{\gamma
[\omega_k-v(-k_1 \sin \delta_t+k_2 \cos \delta_t)]\}^{1/2}\;\;[
\omega_{k^\prime} -\omega_k + v k_2^\prime + v(-k_1 sin \delta_t
+k_2 \cos \delta_t)]}{2\{\omega_{k^\prime}\}^{1/2}}\;\; \tilde{G}
\nonumber \\
\delta(k_1+k_1^\prime)\;\;\delta(k_2+k_2^\prime)\;\;
\delta(k_3+k_3^\prime)\; a^{\mu_t}(k),
\end{eqnarray}
with
\begin{eqnarray}
\tilde{G}=\exp i\{t [-\omega_{k^\prime}-\omega_k + v(-k_2 \sin
\delta_t + k_2 \cos \delta_t)] -r [k_1^\prime + k_1 \cos \delta_t
+ k_2 \sin \delta_t] \}.
\end{eqnarray}
So the expressions in the $\delta-$ functions got very simple
form. \\
\indent Having integrated this expression and taking into
consideration that
\begin{eqnarray}
 \omega_k = \omega_{|-\vec{k}^\prime|}=\omega_{k^\prime},
 \end{eqnarray}
 we obtain:
 \begin{eqnarray}
 \hat{I}_{k^\prime}= \frac{v\;\gamma^{1/2}[\omega_{k^\prime}-v(k_1^\prime
 \sin \delta_t-k_2^\prime \cos \delta_t)]^{1/2}\;[k_2^\prime +
 k_1^\prime \sin \delta_t-k_2^\prime \cos \delta_t]}
 {2 \omega_{k^\prime}^{3/2}} \times \nonumber \\
 \exp\{it [-2 \omega_{k^\prime} + v
 (k_2^\prime \sin \delta_t - k_2^\prime \cos \delta_t)] +
 ir[-k_1^\prime + k_1^\prime \cos \delta_t + k_2^\prime
 \sin \delta_t]\} a^{\mu_t}(-\vec{k}^\prime).
 \end{eqnarray}
 In the same way we could show that
 \begin{eqnarray}
 \int d^3 k
 \beta^{\ast}_{kk^\prime}a^{\mu_t}(k)=
 \hat{I}^{+}_{k^\prime}.
 \end{eqnarray}
 Then because
 \begin{eqnarray}
 <0^{\mu_t}|a^{\mu_t}(-\vec{k}^\prime)|
 a^{+\mu_t}(-\vec{k}^\prime)|0^{\mu_t}>=1,
 \end{eqnarray}
\begin{eqnarray}
<\;0^{\mu_t}\; |N_{k\prime}|\;0^{\mu_t} \;> =
\hat{I}^{+}_{k^\prime}|\hat{I}_{k^\prime}  = \;\frac{v^2 \;\gamma
\; [\omega_{k^\prime} -v (k_1^\prime \sin \delta_t -k_2^\prime
\cos \delta_t)] (k_2^\prime + k_1^\prime \sin \delta_t -k_2^\prime
\cos \delta_t)^2}{4 \omega_{k^\prime}^3}.
\end{eqnarray}
So $< O^{\mu_t}| N_{k^\prime} |O^{\mu_t}> \;\neq 0 \; $. Following
a usual interpretation \cite{birell1982} (3.41), the fact of non
zero value of $\beta_{kk^\prime}$ means that detector rotating in
a vacuum state $|\;0\;>$ observes non zero number of particles of
mode $f_k^{\mu_t}$. The vacuums $|\;0\;>$ and $|\;0^{\mu_t}\;>$
and associated Fock spaces are not unitary equivalent. In the
limit of $v \rightarrow 0$ or $\delta_t \rightarrow 0$
$\beta_{kk^\prime} \rightarrow 0$, and both Fock spaces and
their vacuums agree, what is supposed to be.\\
\indent Specific feature of the Bogolubov transformation found
here is its dependence on time. It is a consequence of the way a
non inertial rotating  detector observes the quantized massless
scalar field in the vacuum state. At each moment the detector uses
that inertial reference frame which is momentarily  at rest
relative to and agrees with it at that moment of time. There is
another example of the time dependent Bogolubov transformation
\cite{grib1980} $\$6.2$. The quantized Fermi field interacting
with  an external classical uniform electric field can be
represented as a free field at any time if a vacuum state at that
moment is redefined correspondingly. So an external classical
field is turned off the same way as the inertial reference system
is switched to the inertial reference frame in our case, by the
redefining a vacuum state of the quantized filed. Our calculation
of the Bogolubov coefficients was greatly motivated by that
result.
\section{The spectrum of the random classical massless scalar field
observed by a rotating  detector.}
\subsection{Periodicity and the correlation function.}
Under the assumption about periodicity  the correlation function
of the massless scalar field at the rotating detector
(\ref{eq:CF_Scal_Field_Continuous})becomes [12.15.04]:
\begin{eqnarray}
\langle \psi_s(\mu_1|A_1^{\mu_1}, t_1^{\mu_1})
\psi_s(\mu_2|A_2^{\mu_2}, t_2^{\mu_2})\rangle = \langle
\psi_s(\lambda_2|A_1^{\lambda_2},t_1^{\lambda_2})
\psi_s(\lambda_2|A_2^{\lambda_2}, t_2^{\lambda_2})\rangle =
\label{eq:CF_Scalar_InTermsOfOneFrame} \\
\langle k_0 \sum_{k_{n_1}}\int dO_1 k_{n_1}^2 f(ck_{n_1})\cos \{
-k_{n_1 x} r(1-\cos\delta) -k_{n_1 y}r \sin \delta- c k_{n_1}t_1
-\theta(\vec{k}_{n_1})\} \times \nonumber \\
\label{eq:CF_Scalar_InTermsOfOneFrame1}
 k_0 \sum_{k_{n_2}} \int dO_2 k_{n_2}^2
f(ck_{n_2})\cos\{-ck_{n_2}t_2 -\theta(\vec{k}_{n_2})\} \rangle.
\end{eqnarray}
On the right side of (\ref{eq:CF_Scalar_InTermsOfOneFrame}) the
coordinates $A_1^{\lambda_2}$, $A_2^{\lambda_2}$ and the times
$t_1^{\lambda_2}$, $t_2^{\lambda_2}$ of both points and the wave
functions are again considered in the same reference frame
$\lambda_2$. In the (\ref{eq:CF_Scalar_InTermsOfOneFrame1}) the
explicit expressions of these wave functions in the $\lambda_2$
reference frame are given. \\
\indent In spherical coordinates the relationship for the massless
scalar field
\begin{eqnarray}
\langle \cos \theta(\vec{k}_{n_1})\cos\theta(\vec{k}_{n_2})
\rangle= \langle \sin
\theta(\vec{k}_{n_1})\sin\theta(\vec{k}_{n_2}) \rangle=
\frac{1}{2} \delta^3(\vec{k}_{n_1}-\vec{k}_{n_2})
\end{eqnarray}
with a discrete spectrum becomes (compare with
(\ref{eq:cosaverage})):
\begin{eqnarray}
\langle\cos[\theta(k_0n_1\bar{k}_1)]\cos[\theta(k_0n_2\bar{k}_2)]\rangle=
\langle\sin[\theta(k_0n_1\bar{k}_1)]\sin[\theta(k_0n_2\bar{k}_2)]\rangle=
\frac{1}{k_0(n_1k_0)^2}\delta_{n_1
n_2}\delta(\bar{k}_1-\bar{k_2}).
\end{eqnarray}
Using this expression the correlation function can be written in
the form:
\begin{eqnarray}
\langle \psi_s(\mu_1|A_1^{\mu_1}, t_1^{\mu_1})
\psi_s(\mu_2|A_2^{\mu_2}, t_2^{\mu_2})\rangle = k_0^2 \sum_n \int
d O (nk_0)^4 f^2(cnk_0)\frac{1}{k_0(n_1k_0)^2} \times \nonumber \\
\cos\{nk_0
[r\bar{k}_x(1-\cos\theta)+r \bar{k}_y \sin\delta -c(t_2-t_1)]\}.
\end{eqnarray}
\indent The Lorentz-invariant spectral function $f_0(ckn_0)$
\cite{boyer1980}, (16), is
\begin{eqnarray}
f^2(cnk_0)=\frac{\hbar c}{2 \pi^2 n k_0}.
\end{eqnarray}
Using (\ref{eq:wavevectors})we can rotate
$(\bar{k}_x,\bar{k}_y,\bar{k}_z)$ to $(\bar{k}_x^{\prime},
\bar{k}_y^{\prime},\bar{k}_z^{\prime })$. Then
$\bar{k}_x(1-\cos\delta)+\bar{k}_y\sin\delta=2
\sin\frac{\delta}{2}\bar{k}_y^{\prime}$ and we arrive to the
expression:
\begin{eqnarray}
\langle \psi_s(\mu_1|A_1^{\mu_1}, t_1^{\mu_1})
\psi_s(\mu_2|A_2^{\mu_2}, t_2^{\mu_2})\rangle = \frac{k_0^2 \hbar
c}{2 \pi^2}\int d o \sum_{n=0}^{\infty}n \cos nF,
\end{eqnarray}
where $do= d\theta d \phi \sin\theta$,  F is defined in
(\ref{eq:F}) and depends on both $\theta$ and $\phi$.
\subsection{ Abel-Plana formula and the temperature of the massless
scalar field observed by a rotating detector.} \indent Abel-Plana
summation formula, we have already discussed above, in this case
is
\begin{eqnarray}
\sum_{n=0}^{\infty}n \cos nF =\int_0^{\infty} dt\; t \cos tF -
\int_0^{\infty} dt \frac{2t \cosh tF }{e^{2 \pi t}-1}
\end{eqnarray}
or
\begin{eqnarray}
\label{eq:planck} \Omega^2 \;\sum_{n=0}^{\infty}n \cos nF
=\int_0^{\infty} d\omega\; \omega \cos \omega \tilde{F} -
\int_0^{\infty} d\omega \frac{2\omega \cosh \omega \tilde{F}
}{e^{\frac{\hbar \omega}{k T_{rot}}}-1},
\end{eqnarray}
where $T_{rot}$ is defined in (\ref{eq:T}). \\
\indent This expression is similar to the expression
[\cite{boyer1980}], (27) for the correlation function of the
detector at rest in Planck's spectrum:
\begin{eqnarray}
\int_0^{\infty}d \omega \omega \coth\frac{\hbar \omega}{2
kT}\cos\omega t =\int_0^{\infty}d \omega \cos \omega t +
\int_0^{\infty}d \omega \frac{2 \omega \cos\omega
t}{e^{\frac{\hbar \omega}{kT}}-1}.
\end{eqnarray}
The likeness between them becomes especially close when $t=0$ and
$\tilde{F}=0$. The appearance of the Planck's factor
$(e^{\frac{\hbar \omega}{kT}}-1)^{-1}$ in (\ref{eq:planck})points
out that the rotating detector in the  massless  scalar zero-point
field observes the same  radiation spectrum as an inertial
observer placed in a thermostat filled up with the radiation at
the temperature $T=T_{rot}$.\\
\\
 {\bf APPENDIX}
\appendix
\section{ $\lambda_\tau$ and $\mu_\tau$ reference frames, Lorentz
transformations, and initial condition.}
 We have already mentioned in the Introduction that global RF's
 $\lambda_\tau$ and
$\mu_\tau$, by definition, are connected by a Lorentz
transformation and agree at any proper time $\tau$, measured by
the detector clock. This initial condition is different from one
used in a usual Lorentz transformation when two inertial systems
agree at the time $t=t^{\prime}=0$ \cite{Landau}.  Let us consider
the connection between these RF's
in detail.\\
\indent We expect that any event ($\vec{x}^\lambda, t^\lambda$) at
$\lambda_\tau$ RF is connected with an event ($\vec{x}^\mu,
t^\mu$) at $\mu_\tau$ RF as \cite{moller}, II.$25^{\prime}$( in
our notations):
\begin{eqnarray}
\vec{x}^\lambda= \vec{x}^\mu + \vec{v}^\mu \frac{\vec{x}^\mu
\vec{v}^\mu}{v^2}(\gamma -1)- \vec{v}^\mu t^\mu \gamma,
           \nonumber       \\
t^\lambda =\gamma t^\mu- \gamma \frac{\vec{v}^\mu
\vec{x}^\mu}{c^2},
                       \label{eq:LorCoor}
\end{eqnarray}
and, because the detector velocity vector
\begin{eqnarray}
(v_1^\mu, v_2^\mu, v_3^\mu)=\vec{v}^\mu= -\vec{v}^\lambda= (0, -v,
0),
\end{eqnarray}
(note that $v_2^{\mu}=-v$ ) the equations have form:
\begin{eqnarray}
x^\lambda_1=x^\mu_1 , & x^{\lambda}_2=(x^{\mu}_2+ v t^\mu)\gamma,
\nonumber \\
x^\lambda_3=x^\mu_3,  &  t^\lambda=(t^\mu +\frac{v}{c^2}x^{\mu}_2
)\gamma
\end{eqnarray}
Under these transformations, \\
$x_2^{\mu}=0$, $t^\mu=\tau$\\
 transform to\\
$x_2^{\lambda}=v \tau \gamma$, $t^\lambda=\tau\gamma$\\
at any detector proper time $\tau$. \\
 The $x_2^{\lambda}$ is not zero, and $\mu_\tau$ and $\lambda_\tau$ do not agree,
 against our expectations. This could mean that
  our assumption that RF's, $\mu_\tau$ and $\lambda_\tau$ agree
at the proper time $\tau$is wrong or the form of the Lorentz
transformation we use here is not correct.\\
 We will now show  that it is the Lorentz transformation that should
 be slightly modified  following the new initial condition. Indeed, the
equations (\ref{eq:LorCoor}) are derived with the assumption
\cite{rohrlich} that
two inertial reference frames agree that is \\
$x_2^\lambda=x_2^\mu=0$ \\
at\\
 $t^\lambda=t^\mu=0$.\\
In our problem, this initial condition is  true, and \\
$t^\lambda=t^\mu=0$, \\
 for only
one pair of the RF's, $\lambda_\tau$ and $\mu_\tau$, when
$\tau=0$. It is false when $\tau
\not=0$. \\
It is easy to see that the modified transformation
\begin{eqnarray}
\label{eq:ModifiedLorentzTransformation}
x_2^{\lambda}=(x_2^{\mu} + v t^{\mu}) \gamma + a_\tau  \\
\nonumber t^{\lambda}=(t^{\mu}+ \frac {v x_2^{\mu}}{c^2}) \gamma
\end{eqnarray}
transform
\begin{eqnarray}
x_2^{\mu}=0, t^{\mu}= \tau
\end{eqnarray}
to
\begin{eqnarray}
x_2^{\lambda}=0, t^{\lambda}= \tau \gamma
\end{eqnarray}
if a constant $a_\tau$, depending on the parameter $\tau$,
 is set to $-v \gamma \tau$.\\
\indent  This modified Lorentz transformation leaves the intervals
invariant ( for simplicity we use here
 2-dimensional interval)
\begin{eqnarray}
(x_2^{\lambda})^2 -c^2(t^{\lambda}- \tau \gamma)^2= (x_2^{\mu})^2
-c^2(t^{\mu}- \tau )^2=0.
\end{eqnarray}
It differs from the usual Lorentz transformation with the initial
condition. For example the [\cite{moller}] successive  Lorentz
transformation has initial condition at $\tau=\tau^{\prime}=0$.
 So we have proved that our assumption in the Introduction, that
the rotating detector can be at the origin of both RF's, $\mu$ and
$\lambda$, at any time $\tau$, are true, and its coordinates in
both RFs are:
 \begin{eqnarray}
A^{\mu}=(x^{\mu}_1,x^{\mu}_1,x^{\mu}_3 )=0, &
A^{\lambda}=(x^{\lambda}_1, x^{\lambda}_2,x^{\lambda}_3)=0
\end{eqnarray}
There is a slight  difference between our definition of RF and the
one used in \cite{irvine}.
  In \cite{irvine}, the $\lambda$ and $\mu$ RF's are defined and
Lorentz transformations are used  in terms of orthogonal tetrades
locally, at a point of the world line of a rotating detector
moving in the non-Minkovskian space-time, with the metrics $g_{\mu
\nu}$. They are not applied to spatial  coordinates and time.
\section{ Hyperbolic motion and Lorentz transformation.}
The concept of an inertial reference frame $I_\tau$ in
\cite{boyer1980}, \cite{boyer1984} is a central part of the
calculation of the correlation function in the case of an
uniformly accelerating point detector. The inertial frame $I_\tau$
is defined by the condition that the point detector  is
instantaneously at rest in $I_\tau$ at the proper time measured by
its clock. Also at time $t_\tau=\tau$ the detector position is at
the origin of $I_\tau$ that is $x_\tau=0$. Then it is assumed that
a Lorentz transformation exists, from $I_\tau$ to the laboratory
coordinate system, which is supposed to transform these two
coordinates to $X_{\star}(\tau)$ and $t_{\star}(\tau)$, defined in
(8) and (9) equations  of \cite{boyer1984}.
\begin{eqnarray}
X_{\star}(\tau)=\frac{c^2}{a}[\cosh(\frac{a\tau}{c})-1] \nonumber \\
t_{\star}=\frac{c}{a}\sinh(\frac{a\tau}{c})
\end{eqnarray}
\indent Correctness of this assumption  has never been proved
explicitly but nevertheless it was used directly in calculations
in (17)and (53) of \cite{boyer1980}.  The coordinates
$X_{\star}(\tau)$ and $t_{\star}(\tau)$ describe the motion of the
detector in the laboratory coordinate system and, as we will show
here, have nothing to do with the coordinates $x_{\tau\star}$ and
$t_{\tau\star}$, which can be obtained as a result of a Lorentz
transformation applied to  $t_\tau=\tau$ and $x_\tau=0$ in
$I_\tau$. Indeed,
 at time $\tau$,  velocity
$v_\tau$ and $\gamma_\tau$ are (3 and 4 in \cite{boyer1980}):
\begin{eqnarray}
v_\tau = c \tanh(\frac {a \tau}{c}),  \nonumber\\
 \gamma_{\tau} =(1- \frac {v^2}{c^2})^{-\frac{1}{2}}=\cosh(\frac{a
 \tau}{c}),
\end{eqnarray}
 and
\begin{eqnarray}
x_{\star\tau}=(x_\tau + v_{\tau} t_\tau)\gamma_{\tau}  \nonumber
\\
t_{\star\tau}= (t_\tau + \frac {v_{\tau} x_\tau}{ c^2})
\gamma_{\tau}.
\end{eqnarray}
If $x_\tau=0$ and $t_\tau=\tau$ then
\begin{eqnarray}
x_{\star\tau}= c \tau \sinh \frac{a \tau}{c}=v_{\tau}\tau\gamma_{\tau},   \nonumber \\
t_{\star\tau}= \tau \cosh\frac{a \tau}{c}=\tau\gamma_{\tau}.
\end{eqnarray}
Obviously
\begin{eqnarray}
x_{\star\tau}\not= X_{\star}(\tau)   \nonumber \\
t_{\star\tau} \not=t_{\star}(\tau)
\end{eqnarray}
 Fortunately, as
we will show,  this assumption does not effect the final results
in the case of a uniformly accelerating detector if in addition to
an infinite set of instantaneous reference frames $I_\tau$ in
\cite{boyer1980}, \cite{boyer1984} we  introduce an infinite set
of inertial reference frames $I_{\tau\star}$, which are at rest
relative to the laboratory coordinate system. This way we have two
reference systems, an inertial one consisting of $I_{\tau\star}$
reference frames and an accelerating one consisting of $I_\tau$
reference frames. The first reference system is similar to  the
$\{\lambda_\tau\}$ reference system and the  second is similar to
the $\{\mu_\tau\}$ reference system described in previous
sections. The locations of the accelerating detector in two
reference frames $I_{\tau_1}$ and $I_{\tau_2}$ are:
\begin{eqnarray}
A_1^{I_{\tau_1}}=(0,0,0), t_1^{I_{\tau_1}}=\tau_1, \nonumber  \\
A_2^{I_{\tau_2}}=(0,0,0), t_2^{I_{\tau_2}}=\tau_2.
\end{eqnarray}
After applying the Lorentz transformation with the initial
condition (\ref{eq:ModifiedLorentzTransformation})  they are
transformed  to their locations in the references frames
$I_{\tau_1\star}$ and $
                   I_{\tau_2\star}$:
\begin{eqnarray}
A_1^{I_{\tau_1\star}}=(0,0,0), t_1^{I_{\tau_1\star}}=\gamma_{\tau_1}\tau_1, \nonumber  \\
A_2^{I_{\tau_2}\star}=(0,0,0),t_2^{I_{\tau_2}\star}=\gamma_{\tau_1}\tau_2.
\end{eqnarray}
The references frames $I_{\tau_1\star}$ and $I_{\tau_2\star}$ have
the same axis directions but are shifted against each other by the
distance

\begin{eqnarray}
X_{\star}(\tau_2)-X_{\star}(\tau_1)=\frac{c^2}{a}(\cosh(\frac{a\tau_2}{c})-\cosh(\frac{a\tau_1}{c}))
\end{eqnarray}

\end{document}